\documentclass[aip,jcp,preprint,longbibliography]{revtex4-1}
\usepackage{color}
\usepackage{amsmath}
\usepackage{amssymb}
\usepackage{graphicx}
\usepackage[T1, OT1]{fontenc}
\usepackage{hyperref}

\newcommand{\DGcnf}{ \Delta G^\mathrm{cnf}}
\newcommand{\on}{\overline{n}}

\begin{document}

\title{Melting transition in lipid vesicles functionalised by mobile DNA linkers\footnote{$^{*,\dag}$These authors contributed equally to this work; $^{\dag}$corresponding authors: ld389@cam.ac.uk, bmognett@ulb.ac.be}}

 \author{S.\ J.\ Bachmann$^*$}
 \affiliation{Universit\'e Libre de Bruxelles (ULB), Interdisciplinary Center for Nonlinear Phenomena and Complex Systems \& Service de Physique des Syst\`emes Complexes et M\'ecanique Statistique, Campus Plaine, CP 231,
Blvd du Triomphe, B-1050 Brussels, Belgium.}
 \author{J.\ Kotar$^*$}
 \affiliation{Biological and Soft Systems, Cavendish Laboratory, University of Cambridge, JJ Thomson Avenue, Cambridge, CB3 0HE, United Kingdom}
 \author{L.\ Parolini}
 \affiliation{Biological and Soft Systems, Cavendish Laboratory, University of Cambridge, JJ Thomson Avenue, Cambridge, CB3 0HE, United Kingdom}
 \author{A.\ {\v S}ari{\'c}}
 \affiliation{Department of Chemistry, University of Cambridge, Cambridge CB2 1EW, United Kingdom; Department of Physics and Astronomy, Institute for the Physics of Living Systems, University College London, WC1E 6BT, United Kingdom. }
 \author{P.\ Cicuta}
 \affiliation{Biological and Soft Systems, Cavendish Laboratory, University of Cambridge, JJ Thomson Avenue, Cambridge, CB3 0HE, United Kingdom}
 \author{L.\ Di Michele$^\dag$}
 \affiliation{Biological and Soft Systems, Cavendish Laboratory, University of Cambridge, JJ Thomson Avenue, Cambridge, CB3 0HE, United Kingdom}
 \author{B.\ M.\ Mognetti$^\dag$}
 \affiliation{Universit\'e Libre de Bruxelles (ULB), Interdisciplinary Center for Nonlinear Phenomena and Complex Systems \& Service de Physique des Syst\`emes Complexes et M\'ecanique Statistique, Campus Plaine, CP 231,
Blvd du Triomphe, B-1050 Brussels, Belgium.}

\begin{abstract}
We study  phase behaviours of lipid--bilayer vesicles functionalised by ligand--receptor complexes made of synthetic DNA by introducing a modelling framework and a dedicated experimental platform. 
In particular, we perform Monte Carlo simulations that combine a coarse grained description of the lipid bilayer with state of  art analytical models for multivalent ligand--receptor interactions. 
Using density of state calculations, we derive the partition function in pairs of vesicles and compute the number of ligand--receptor bonds as a function of temperature. 
Numerical results are compared to microscopy and fluorimetry experiments on Large Unilamellar Vesicles decorated by DNA linkers carrying complementary overhangs. 
 We find that vesicle aggregation is suppressed when the total number of linkers falls below a threshold value. 
Within the model proposed here, this is due to the higher configurational costs required to form inter--vesicle \emph{bridges} as compared to intra-vesicle {\em loops}, which are in turn related to membrane deformability. 
Our findings and our numerical/experimental methodologies are applicable to the rational design of liposomes used as functional materials
and drug delivery applications, as well as to study inter-membrane interactions in living systems, such as 
cell adhesion. 
\end{abstract}

\maketitle
\section{Introduction}

Ligand-receptor interactions play a crucial role in a large variety of biological processes, including cell adhesion and signalling.\cite{GayReview2014}
In biology, the selective nature of ligand-receptor interactions enables functional behaviours and responsiveness to environmental stimuli, which can be replicated in biomimetic materials where the interactions between the unit components are mediated by supramolecular ligand--receptor complexes.\cite{sackmann2014physics}
The eminent example is represented by systems of DNA-functionalised colloidal units,\cite{mirkin,alivisatos,lorenzo,seeman-mirkin-review} where the artificial DNA linkers can be designed to control phase behaviour\cite{crystal-mirkin,crystal-gang,pine,halverson2013dna,stefano-nature,rogers-manoharan,mcginley2013assembling}
and self-assembly kinetics,\cite{miriam-nature,chaikin-subdiffusion,rogers2013kinetics,halverson2016,parolini2016,wang2015crystallization} as well as to engineer biological probes.\cite{francisco-pnas,licata2008kinetic,taton,nam2003nanoparticle}\\
A complete understanding of the complex emerging phenomena observed in multivalent interactions is only possible through a combination of experiments and numerical/theoretical analysis.
Modelling multivalent interactions \cite{mammen1998polyvalent}  is however a challenging task, as it requires the calculation of ensemble averages over the many possible configurations of the supramolecular linker complexes. \cite{kitov2003nature,francisco-pnas} 
Analytical methods capable of capturing the resulting entropic effects have been recently developed.\cite{melting-theory1,melting-theory2,crocker-pnas,patrick-jcp,stefano-jcp,di2016communication,tito2016communication,xu2016simple,DeGernier2014,shimobayashi2015direct} 
Such theories have been utilised to calculate effective potentials in DNA mediated interactions between solid\cite{melting-theory1,melting-theory2,crocker-pnas,patrick-jcp,stefano-jcp,di2016communication} and deformable paricles,\cite{parolini2014thermal,shimobayashi2015direct,feng2013specificity,pontani2012biomimetic,stefano-prl}
as well as to design superselective probes.\cite{kitov2003nature,francisco-pnas,xu2016simple,DeGernier2014,tito2016communication,shimobayashi2015direct,dubacheva2015designing} 
Analytical models have however limited applicability to those biologically and technologically relevant scenarios where non-specific contributions significantly affect the interactions. Effects neglected by analytical models include steric repulsion between the linkers and the deformability of the interacting surfaces. The latter is a particularly crucial aspect when dealing with multivalent interactions between soft substrates such as biological membranes.
Numerical approaches could provide a faithful description of specific and non-specific effects in multivalent interactions. However, in view of the large interval of relevant lenghtscales, from the molecular scale of ligands/receptors to the micron scale of the interacting units, even coarse-grained models are unsuitable to simulate phase behaviour and material properties in ensambles of micron-sized multivalent objects.\cite{angioletti2016theory}  Overcoming the limitation of purely analytical and purely numerical approaches is a critical step towards the development of truly predictive theoretical methods to aid the design of synthetic multivalent materials and improve our understanding of emergent behaviours in multivalent biological systems. A suitable approach should be capable of describing simultaneously the behaviour of the individual ligand/receptor pairs and the global phase behaviour of the systems.\\
The validation of such multiscale theoretical framework requires dedicated experiments in which the global phase behaviour can be disentangled from that of the individual ligand/receptor pairs, and the effect of the latter on the former can be assessed as a function of the number and adhesive strength of the linkers.\\
In this work we investigate the self-assembly behaviour of a biomimetic system of DNA-functionalised lipid vesicles,\cite{parolini2014thermal,shimobayashi2015direct,Hadorn2013,banga2014liposomal,beales2007,parolini2016} where artificial DNA linkers play the role of ligands/receptors and membrane deformability can potentially affect the resulting multivalent interactions.  DNA linkers can freely diffuse on the surface of the vesicles and form either intra-vesicle \emph{loops} or inter-vesicle \emph{bridges}, the latter being responsible for attractive interactions and driving vesicle aggregation. We study the response of the system to temperature changes, and clarify how the aggregation/melting transition of the liposomes is affected by the competition between loop and bridge formation and the non-selective free energy contributions related in turn to membrane deformability.\\
We propose a new "hybrid" framework to calculate
the free energy of the interactions between such vesicles, that combines state-of-art analytical theories developed for solid particles \cite{patrick-jcp,stefano-jcp,di2016communication,parolini2014thermal,shimobayashi2015direct,feng2013specificity,pontani2012biomimetic,stefano-prl} with Monte Carlo simulations that account for configurational costs related to membrane deformability.\\
By exploiting a fully automated and programmable platform, we perform experiments that for the first time are capable of simultaneously monitoring the self-assembly state of the liposomes and the bound/unbound state of the DNA linkers through a combination of fluorescence microscopy and F\"{o}rster Resonance Energy Transfer (FRET) measurements.\\
In systems where the number of linkers per vesicle $N$ is low, simulations are capable of quantitatively 
replicating the response to temperature changes observed in experiments, including the aggregation/melting temperature of 
 Large Unilamellar Vesicle (LUV) clusters and its correlation with the temperature-dependent fraction of formed DNA bonds. 
Such agreement is lost at high $N$, most likely due to the effects of linker-linker and linker-vesicle steric interactions, neglected by our current model.
Our numerical results confirm the importance and the nature of the previously hypothesised entropic effects on the hybridisation free energy of surface-tethered linkers.\cite{feng2013specificity,parolini2014thermal,stefano-prl,miriam} In cases where linkers can diffuse on the surface of the substrates, such as DNA-tethers on lipid vesicles or ligands/receptors on cells,\cite{smith2008force} these entropic costs include the loss of translational freedom following the formation of a bond. Particularly severe are the effects experienced by linkers forming inter-vesicle bridges, which end up confined within the relatively small \emph{adhesion patch} between the vesicles. With the present numerical method these contributions can be directly evaluated based on the observed size of the adhesion patch, and then compared to the overall configurational free energy costs evaluated \emph{via} density of states calculations.\cite{parolini2014thermal,stefano-prl}\\
Experiments and simulations demonstrate that these repulsive free energy contributions, combined with the competition between loop and bridge formation, have a substantial effect on self-assembly behaviour of DNA-functionalised vesicles: a \emph{minimum number of linkers $N_\mathrm{dim}$} is required to stabilise adhesion.\cite{stefano-nature,rogers-manoharan} If fewer linkers are present vesicles do not aggregate.\cite{stefano-nature,rogers-manoharan}
The fair agreement between experimental and predicted value of the threshold number of linkers validates our methods as an useful tool to design biomimetic self-assembling systems  featuring complex functionalities. \\
%, as well as to investigate cell adhesion.\\

%The remainder
The paper is structured as follows. Sec. \ref{sec:experimental}  reports on experimental methods. In Sec.~\ref{sec:model} we present our  modelling framework. In particular in Sec.\ \ref{sec:multivalent} and Sec.\ \ref{sec:freeenergy} we present respectively the analytical and numerical methods. The latter section illustrates how the two are combined.
In  Sec. \ref{sec:res} we discuss the simulations' outcomes for a large set of different system parameters (listed in Sec.\ \ref{sec:ana}) and perform a detailed comparison with experimental results.\\

%%%%%%%%%%%%%%%%%%%%%%%%%%%%%%%%%%%%%%%%%%
%%%%%%%%%% FIGURE 1 %%%%%%%%%%%%%%%%%%%%%%%%%%
%%%%%%%%%%%%%%%%%%%%%%%%%%%%%%%%%%%%%%%%%%

\begin{figure*}[ht!]
\includegraphics[width=18cm]{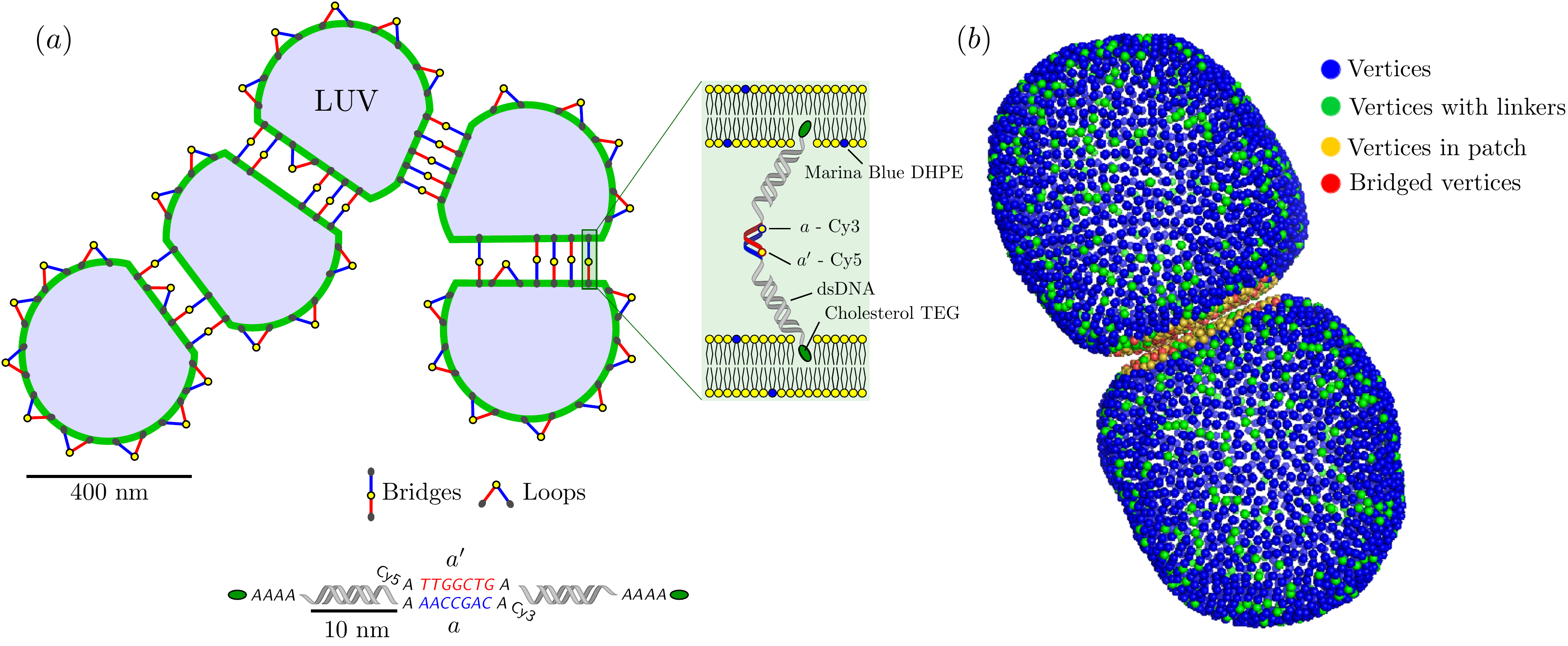} 
\caption{(a) Experimental system. We consider LUV functionalised by two species of DNA linkers carrying complementary sticky ends $a$ and $a'$ and a cholesterol anchor that partitions inside the lipid bilayer. Sticky ends $a$ and $a'$ are labelled with fluorophores $Cy3$ and $Cy5$ as shown in the schematics. Two kind of bonds are possible: intra-vesicle loops and inter-vesicle bridges, the latter drive self-assembly of the suspension.
(b) Simulation system. Lipid bilayers are modelled as triangulated meshes whose vertices are hard spheres (blue elements). Some of the vertices carry a reactive linker (green elements) that can eventually bind a complementary linker when closer than a distance $L$.  Linkers forming a bond are labelled in red. In this snapshots loop formation is not allowed. 
The yellow elements highlight the vertices belonging to the adhesion patch between the vesicles.}
\label{fig:figure1}
\end{figure*}

%%%%%%%%%%%%%%%%%%%%%%%%%%%%%%%%%%%%%%%%%%
%%%%%%%%%% EXPERIMENTS %%%%%%%%%%%%%%%%%%%%%%%
%%%%%%%%%%%%%%%%%%%%%%%%%%%%%%%%%%%%%%%%%%

\section{  Experimental Methods }
\label{sec:experimental}
In this section we present our sample preparation protocols and experimental setup designed to provide a complete characterisation of the self-assembly behaviour of DNA-functionalised LUVs. In particular we can simultaneously monitor temperature dependent vesicle aggregation and the binding/unbinding state of the membrane-anchored DNA linkers in multiple samples using fully programmable and automated microscopy/fluorimetry apparatus. Experimental information can then be directly compared to numerical results to validate the assumptions and highlight the limitations of the numerical/theoretical framework described in Sec.~\ref{sec:model}.

\subsection{ Experimental system}

As sketched in Fig.~\ref{fig:figure1}~(\emph{a}), LUVs with diameter $D_\mathrm{exp}\approx 400$~nm are functionalised by hydrophobised DNA linkers featuring a cholesterol anchor partitioning within the lipid bilayer, a rigid double-stranded DNA (dsDNA) spacer of length $\ell \approx 10$~nm and a single-stranded DNA (ssDNA) \emph{sticky end}. Membranes are in a fluid state, which enables cholesterol anchors to freely diffuse onto the vesicles' surface. On average, half of the overall $2N$ linkers present on each vesicle carry a sticky end $a$, the second half carry the complementary sequence $a'$. One linker type can bind to the other forming either inter-vesicle \emph{bridges} or intra-vesicle \emph{loops}, the latter being possible because each vesicle carries both types of linkers. Bridges are responsible for inter-vesicle adhesive forces that ultimately lead to aggregation of the LUVs. Sticky ends $a$ and $a'$ are labelled with fluorophores Cy3 and Cy5 respectively, so that F\"{o}rster Resonance Energy Transfer (FRET) can be used to assess bond formation.

\subsection{ Sample preparation}
\label{sec:prep}
LUVs are prepared from Dioleoyl-sn-glycero-3-phosphocholine (DOPC, Avanti Polar Lipids) doped with 0.8 molar percent  Marina Blue 1,2-Dihexadecanoyl-sn-Glycero-3-Phosphoethanolamine (DHPE, Life Technologies) for fluorescent imaging. Vesicles are prepared by extrusion in a 300 mM sucrose solution using a Mini Extruder (Avanti Polar Lipids) equipped with polycarbonate track-etched membranes with 400 nm pores (Whatman), as detailed in Ref.\ \cite{parolini2016}.\\
Hydrophobised DNA linkers are pre-hybridised from the ssDNA\\
(i) 5' -- \textbf{CGT GCG CTG GCG TCT GAA AGT CGA TTG CG} \emph{AAAA} --3'  [Cholesterol TEG] \\
(ii) 5' -- \textbf{GC GAA TCG ACT TTC AGA CGC CAG CGC ACG} \emph{A} [Sticky End] \emph{A} -- 3' Cy3/Cy5,\\
where bases marked in bold belong to the dsDNA rigid spacer and unpaired \emph{A} bases highlighted in italic are included to provide flexibility.
Hybridisation is carried out in TE buffer (10 mM Tris, 1mM EDTA, Sigma) with 100 mM NaCl as detailed in Ref. \cite{shimobayashi2015direct}.\\
Samples are prepared by mixing a 10~$\mu$l extruded vesicle solution with 90~$\mu$l iso-osmolar solution containing TE buffer, 87 mM glucose, 100 mM NaCl and a variable concentration of pre-assembled DNA linkers spanning from 12.7 $\mu$M to 0.25 $\mu$M, equally divided between $a$ and $a'$ linkers. The nominal number of DNA linkers \emph{per} vesicles is calculated assuming that all linkers partition into the bilayer,~\cite{pfeiffer2004} that all the processed lipids form unilamellar vesicles with diameter of $400$~nm, and that each lipid molecule contributes with 70~\AA$^2$ to the bilayer area.~\cite{Kucerka2006} This estimate leads to a number of $a$ and $a'$ linkers \emph{per} vesicle between 71 $\le N \le$ 3555. The assumption that all linkers partition onto the bilayer may breakup at high DNA density due to surface saturation, as discussed in Sec. \ref{sec:res}.\\
For thermal processing and imaging, samples are injected into borosilicate glass capillaries (0.2$\times$4 mm$^2$ rectangular inner section, CM Scientific), protected with a droplet of mineral oil at both ends, and permanently sealed with epoxy glue (Araldite).

\subsection{ Temperature cycling and imaging }
\label{sec:thermal}
Imaging and fluorescence spectroscopy are carried out on a fully automated Nikon Eclipse Ti-E inverted epifluorescence microscope equipped with a Nikon PLAN APO 20$\times$ 0.75 NA dry objective and a Point Grey Research Grasshopper-3 GS3-U3-23S6M-C camera. Temperature is controlled with a tailor made Peltier stage hosting 7 capillaries with samples at different linker concentration.
Temperature is repeatedly ramped up and down between 80$^\circ$C (or 90$^\circ$C) and 0$^\circ$C in steps of 1$^\circ$C and a rate of $0.5$ $^\circ$C/min, and measured with a precision $<0.1^\circ$C by a thermocouple located in close proximity of the capillaries.
 
At each temperature step, the motorised stage is moved to sequentially image all samples, recording snapshots with different fluorescence excitation wavelength  and emission windows.  For these different combinations of monochromatic LED sources (Philips Lumiled LUXEON UV, LUEXON Z), quad band flourescence filter set (Semrock LED-DA/FI/TR/Cy5-A) and exchangeable motorized emission filters (Semrock FF01-600/37-25, FF01-731/137-25) are used. 
Specifically: Marina Blue  (quad band filter set, excitation LED LHUV-0380-0200 at 380 nm); Cy3-Cy3 (quad band filter set, emission filter FF01-600/37-25, excitation LED LXZ1-PM01 at 530 nm); Cy3-Cy5 (quad band filter set, emission filter FF01-731/137-25, excitation LED LXZ1-PM01 at 530 nm); Cy5-Cy5 (quad band filter set, emission filter FF01-731/137-25, excitation LED LXZ1-PD01 at 627 nm). A single snapshot for each fluorescence channel is recorded on each sample, imaging an area of  563$\times$352 $\mu$m$^2$ at a fixed height of a $\approx$ 8 $\mu$m from the bottom of the capillary.
Dark frames where no excitation is used are acquired between each fluorescence image and used for background subtraction. 
All steps of the experiment are fully automated. A Perfect Focusing System (Nikon) enables correction for vertical drift.

\subsection{ Image analysis  }
\label{sec:imageanalysis}
Images collected on the Marina Blue channel are used for structural characterisation of vesicle aggregates, carried out as described in Ref. \cite{parolini2016} Briefly, Fourier analysis of the images enables the evaluation of a 2D projection of the samples' structure factor $S(q)$, where $q$ is the spatial frequency. The first moment of $S(q)$ is then measured as an indicator of the aggregation state of the samples, going from high values for monomeric vesicles towards low values when aggregation takes place and $S(q)$ develops a strong peak at low-$q$.\\
Images in the Cy3-Cy3, Cy3-Cy5 and Cy5-Cy5 channels are used to assess the efficiency of FRET between donor and acceptor fluorophores attached to $a$ and $a'$ sticky ends respectively. The average intensities of each channel ${\cal I}_\mathrm{Cy3-Cy3}$, ${\cal I}_\mathrm{Cy3-Cy5}$, and ${\cal I}_\mathrm{Cy5-Cy5}$ are extracted from the images, and used to evaluate the (ratio)$_\mathrm{A}$
\begin{equation}\label{ratioA}
\mathrm{(ratio)}_\mathrm{A} = \frac{{\cal I}_\mathrm{Cy3-Cy5} - \alpha~{\cal I}_\mathrm{Cy3-Cy3}}{{\cal I}_\mathrm{Cy5-Cy5}},
\end{equation}
where $\alpha = \tilde{{\cal I}}_\mathrm{Cy3-Cy5} / \tilde{{\cal I}}_\mathrm{Cy3-Cy3}=0.16$, and  the intensities $\tilde{{\cal I}}_\mathrm{Cy3-Cy5}$ and $\tilde{{\cal I}}_\mathrm{Cy3-Cy3 }$ are measured in a reference sample that only contains Cy3 fluorophores. The (ratio)$_\mathrm{A}$ is linearly dependent on the FRET efficiency.~\cite{Clegg1993} The F\"{o}rster radius of the Cy3-Cy5 pair is $\approx 5-6$\,nm,~\cite{Yuan2007} therefore high FRET efficiency is expected in bound $a-a'$ pairs, where donor and acceptor are kept at $\approx 2-3$\,nm from each other. The probability of FRET between unbound linkers is comparatively small, although not fully negligible in samples with the high DNA coverage, where the average distance between unbound $a-a'$ linkers goes down to $\approx 11.9$~nm. As a function of temperature, $\mathrm{(ratio)}_\mathrm{A}$ describes a sigmoidal curve from which the fraction of hybridised linkers $\phi(T)$ can be extracted as
\begin{equation}
\phi(T) = \frac{\mathrm{(ratio)}_\mathrm{A}(T) - B_\mathrm{H}(T)}{B_\mathrm{L}(T) - B_\mathrm{H}(T)},
\end{equation} 
where $B_H$ and $B_L$ are linear fits of the high- and low- temperature plateaus of the sigmoidal curve. See discussion in Sec. \ref{sec:res}.

%%%%%%%%%%%%%%%%%%%%%%%%%%%%%%%%%%%%%%%%%%
%%%%%%%%%% MODELLING %% %%%%%%%%%%%%%%%%%%%%%%%
%%%%%%%%%%%%%%%%%%%%%%%%%%%%%%%%%%%%%%%%%%

\section{  Modelling framework}  
\label{sec:model}

\subsection{ Interacting Large Unilamellar Vesicles (LUV)}
Vesicles are modelled as triangulated surfaces [see Fig.\ \ref{fig:figure1} ($b$)] as described previously.\cite{vsaric2012fluid,vsaric2013self}
Following Refs.~\cite{Nelson86,Ho_EPL,Ho_PRA,kroll97} 
the vertices of the mesh are represented by hard spheres of diameter $\sigma$, taken as the unit length in our simulations. 
With the exception of the case described in Sec.\ \ref{sec:Q1}, in this work we always simulate pairs of interacting vesicles.
The Hamiltonian of two vesicles, each with $N_\mathrm{v}$ vertices is then given by
\begin{eqnarray}
{\cal H}_\mathrm{ves} = \kappa  \sum_{\langle \alpha , \beta\rangle } \, {\bf n}_\alpha \cdot {\bf n}_\beta + \sum_{\langle i,j\rangle} v_\mathrm{bnd}(r_{ij}) + \sum_{i<j} v_\mathrm{excl} (r_{ij}) \, .
\label{eq:ves}
\end{eqnarray}
where $\langle \alpha,\beta\rangle$ and $\langle i, j \rangle$ denote neighbouring triangles and vertices respectively, with $i,j\leq 2N_\mathrm{v}$.
In Eq.\ \ref{eq:ves} $\kappa $  is the bending rigidity, ${\bf n}_\alpha$ is the outward normal of the triangle labelled by $\alpha$, 
 and $v_\mathrm{bnd}(r_{ij})$ is an infinite square-well potential that constraints two neighbouring vertices to be within a  distance $r_\mathrm{cut}=\sqrt{3}\sigma$.
$v_\mathrm{excl}(r_{ij})$ is a hard core repulsion that constraints vertices $i$ and $j$ to stay at distance $r_{ij} > \sigma$. Note that $v_\mathrm{excl}$ acts also between vertices belonging to different vesicles.
%thus for systems with multiple vesicles
%the second sum in the right-hand side term of Eq.~\ref{eq:ves} should be extended to vertices belonging to different vesicles (see ${\cal H}_\mathrm{\mathrm{excl},\alpha,\beta}$ in Eq.\ \ref{eq:TotHAm}).
\\
We distribute 2$N$ implicit linkers, $N$ of type $a$ and $N$ of type $a'$, over the $N_\mathrm{v}$ vertices of each vesicle in a way that no more than one linker  is allowed on the same vertex ($2N<N_\mathrm{v}$). 
Two free  complementary linkers can react if their distance is smaller than $L$.
In the case of DNA linkers, $L$ is equal to twice the length $\ell$ of the spacers connecting the sticky ends to the cholesterol anchors.
Accordingly, the Hamiltonian associated to linker-linker interactions is given by
\begin{eqnarray}
{\cal H}_\mathrm{link}= \sum_{m=1}^{2  N } \sum_{n=1}^{2  N }\Delta G_{m,n} \epsilon_{m,n}
\label{eq:ham-selec}
\end{eqnarray}
where $m$ and $n$ run over all linkers of type $a$ and $a'$ respectively and 
$\epsilon_{m,n}=1$ if  linkers $m$ and $n$ are bound, $0$ otherwise. 
The hybridisation free energy $\Delta G_{m,n}$ is equal to  $\Delta G_\mathrm{L}$ if linkers $m$ and $n$ belong to the same vesicle and can form a loop, or to $\Delta G_\mathrm{B}$ if the linkers are on different vesicles and can form a bridge. The form of $\Delta G_\mathrm{L}$ and $\Delta G_\mathrm{B}$ is discussed in Sec.\ \ref{sec:multivalent}.
The overall Hamiltonian is then given by 
\begin{equation}
{\cal H}= {\cal H}_{\mathrm{ves}} + {\cal H}_\mathrm{link}.
\label{eq:TotHAm}
\end{equation}
% where the indices $\alpha$ and $\beta$ run over all the vesicles in the system and ${\cal H}_\mathrm{\mathrm{excl},\alpha,\beta}$ accounts for the excluded volume terms between vertices of vesicle $\alpha$ and $\beta$. 
If only  ${\cal H}_\mathrm{link}$ is considered, analytical models used in our previous studies\cite{parolini2014thermal,shimobayashi2015direct} enable the calculation of the parition function of two interacting membranes, neglecting non-specific membrane membrane interactions and configurational contributions associated to membrane deformation. These calculations adapted to the present system are given in Sec.\ \ref{sec:multivalent}. The hybrid numerical/analytical framework discussed in Sec.\ \ref{sec:freeenergy} combines analytical calculations with Monte Carlo methods to sample the non-specific contributions described by ${\cal H}_{\mathrm{ves}} $ (see Eq.\  \ref{eq:TotHAm}) to the overall partition function $Z$, which is then used in Sec.\ \ref{sec:Pdim} (see Eq.\ \ref{eq:Z_tot}) to investigate vesicle dimerisation.

%In previous studies\cite{parolini2014thermal,shimobayashi2015direct} we only considered  ${\cal H}_\mathrm{link}$ when calculating partition functions of {\cb interacting} functionalised membranes. These calculations are presented in Sec.\ \ref{sec:multivalent}, adapted to the present system.
%In Sec.\ \ref{sec:freeenergy} we introduce the simulation framework that will be used to sample the contribution given by ${\cal H}_{\mathrm{ves}} $ (see Eq.\  \ref{eq:TotHAm}) to the partition function.
%
%Sec.\ \ref{sec:freeenergy} also illustrate how the analytic and numerical calculations are merged into the final partition function $Z$ that will be used in Sec.\ \ref{sec:Pdim} (see Eq.\ \ref{eq:Z_tot}) to calculate dimerisation transitions.  

\subsection{ Analytical modelling of multivalent interactions }
\label{sec:multivalent}

Interactions between multivalent objects are strengthened by the combinatorial entropic contributions accounting for the many different ways of forming a given number of bonds starting from a set of linkers. 
The experimental system features identical vesicles functionalised by an equal number $N$ of two complementary linkers, which can therefore form  both inter-vesicle bridges and intra-vesicle loops [see Fig.\ \ref{fig:figure1}~($a$)]. In this scenario, 
the contribution of ${\cal H}_\mathrm{link}$ (Eq.\ \ref{eq:ham-selec}) 
to the partition function of two vesicles linked by $n_\mathrm{B}$ bridges, taking two non interacting vesicles as reference, becomes\cite{parolini2014thermal} 
 
\begin{eqnarray}
\Omega_{ \mathrm{L}, \mathrm{B}}(N, n_\mathrm{B}) &=&  {1\over \Omega_\mathrm{L}(N,N)^2} 
\sum_{i=0}^{\mathrm{min}[N,n_\mathrm{B}]} \Omega_\mathrm{B}( N, i) \Omega_B( N, n_\mathrm{B}-i) \cdot 
\nonumber \\
&&  \qquad\cdot  \Omega_\mathrm{L}(N-i,N-n_\mathrm{B}+i )^2 \, ,
\label{eq:multivalency-b+l}
\end{eqnarray}
where
\begin{eqnarray}
 \Omega_\mathrm{L}(K,M) &=& \sum_{i=0}^{\mathrm{min}[K,M]} {K \choose i}  {M \choose i} i! e^{-\beta  \Delta G_\mathrm{L} \cdot i } \, , 
 \label{eq:multivalency-l}
 \\
 \Omega_\mathrm{B} (N, n_\mathrm{B}) &=& {N \choose n_\mathrm{B} }^2 n_\mathrm{B}! \, e^{- \beta \Delta G_\mathrm{B} \cdot n_\mathrm{B} } \, .
\label{eq:multivalency-b}
\end{eqnarray} 
In Eqs.\ \ref{eq:multivalency-b+l} and \ref{eq:multivalency-l} we sum over all the possible number of loops since, consistently with neglecting steric interaction between linkers, we assume that loops are not affected by the conformation of the vesicles.
In Eq.~\ref{eq:multivalency-b+l} $\Omega_\mathrm{L}(N,N)$ is the ``linker'' partition function of an isolated vesicle featuring only loops.  
$\Delta G_\mathrm{L}$ and $\Delta G_\mathrm{B}$ are defined as the free energies for loop and bridge formation.
For system of mobile linkers $\Delta G_\mathrm{B}$ includes the dimerisation free energy of the reactive groups of the linkers when free in solution, indicated as $\Delta G^0$, and a term $\Delta G^\mathrm{rot}$ accounting for the loss of rotational freedom following the binding of two linkers.
In this study we focus on the case of rod-like dsDNA linkers tipped by reactive ssDNA sticky ends, therefore $\Delta G^0$ is simply the hybridisation free energy of the sticky ends as obtained using nearest-neighbour rules for the sequences reported in Fig.\ \ref{fig:figure1}~($a$).\cite{santalucia,santalucia2004thermodynamics,dirks2007thermodynamic} 
 We estimate the rotational term as $\Delta G^\mathrm{rot}= - k_\mathrm{B} T \log \left[1/(\rho_\ominus L^3) \right]$, where $\rho_\ominus=$1M is the standard concentration.\cite{parolini2014thermal,shimobayashi2015direct}
More accurate configurational terms can be estimated accounting for variations in the distance between the two binding linkers. We leave such refinement to future investigations in which linkers will be simulated as explicit objects.
 Differently from $\Delta G_\mathrm{B}$, $\Delta G_\mathrm{L}$ should also include an entropic cost $\Delta G_\mathrm{L}^\mathrm{confining} = -T \Delta S_\mathrm{L}^\mathrm{confining}$ of confining 
  the otherwise diffusive linkers to within a short distance from each other.\cite{parolini2014thermal,shimobayashi2015direct,feng2013specificity,pontani2012biomimetic,stefano-prl}  
Including  an analogous term $\Delta G_\mathrm{B}^\mathrm{confining}$ in $\Delta G_\mathrm{B}$ is not necessary, as confinement entropy is already accounted for by our simulation procedure that explicitly enables surface mobility in bridge-forming linkers (see Sec.\ \ref{sec:sim}).
We estimate $\Delta G_\mathrm{L}^\mathrm{confining}= - k_\mathrm{B}  T \log {\left(L^2 / A\right)}$, where $A$ is the area of a vesicle.\cite{parolini2014thermal,shimobayashi2015direct}  In summary, we obtain
\begin{eqnarray}
\Delta G_\mathrm{L} &=& \Delta G^0 +  \Delta G^\mathrm{rot}  + \Delta G_\mathrm{L}^\mathrm{confining}  = \Delta G^0 - k_B T \log { 1 \over \rho_\ominus L A }
\label{eq:deltagl}
\nonumber \\
\Delta G_\mathrm{B} &=& \Delta G^0 +  \Delta G^\mathrm{rot}   = \Delta G^0 - k_B T \log { 1 \over \rho_\ominus L^3 } \, .
\label{eq:deltagl}
\end{eqnarray}
Recently  it has been reported that the molecular roughness of the bilayer can alter the affinity between complementary linkers. \cite{hu2013binding} When properly parametrised,\cite{xu2015binding} such effects can be included into the definition of $\Delta G_0$.\\
In this work we combine multivalent partition functions, like the one in Eqs.\ \ref{eq:multivalency-b+l} or \ref{eq:multivalency-b}, with Monte Carlo estimates of the configurational free energy costs of pairs of vesicles linked by $n_\mathrm{B}$ inter-vesicle bridges. 
 In our model such costs only depend on the number of bridges $n_\mathrm{B}$. In particular, the interaction free energy does not depend on the number of formed loops (see Sec.\ \ref{sec:DGcnf}). For computational efficiency we therefore use a simplified system where only bridges are possible, which could be realised experimentally using two families of vesicles each carrying only one of the two complementary linkers. In this case the contribution to the partition function due to ${\cal H}_\mathrm{link}$ for vesicles with $n_B$ bridges and $N$ linkers ({\em per} type) is simply given by $\Omega_\mathrm{B}(N,n_\mathrm{B})$ (Eq.\ \ref{eq:multivalency-b}). 
In Sec.\ \ref{sec:freeenergy} we use $\Omega_\dag$ (e.g.\ $\dag=\mathrm{B}$ or $\dag=\mathrm{B,L}$) to tag the contribution to the partition function due to the selective part of a generic multivalent system.

\begin{figure}[t!]
\includegraphics[width=9cm]{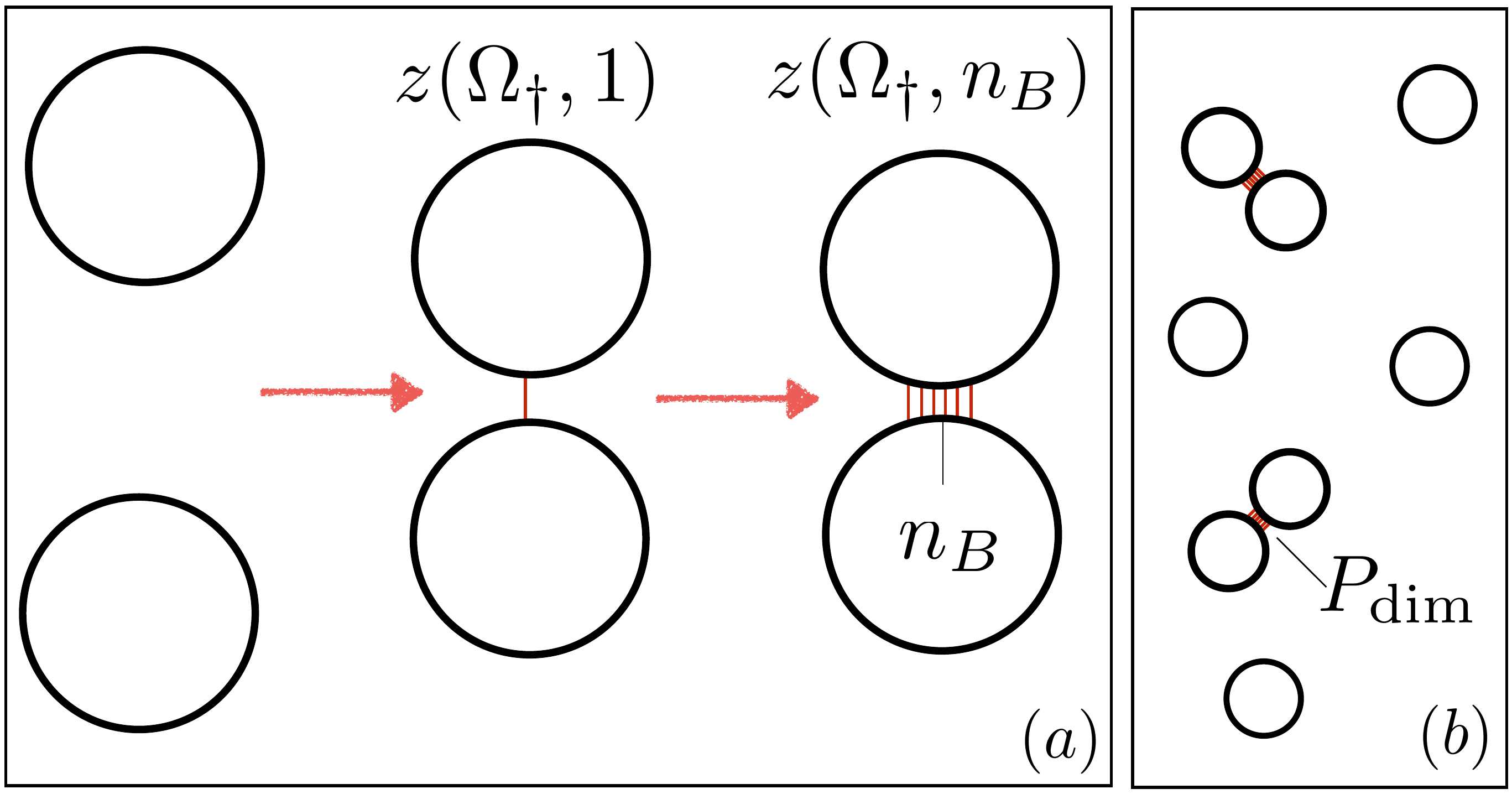} 
\caption{ Thermodynamic pathway for the determination of vesicle dimerisation partition function. $(a)$ The configurational costs entering the partition function of dimers are calculated by first sampling the conformation of two vesicles that can feature at least one bridge (Sec.\ \ref{sec:Q1}), and then by sequentially adding more bridges (Sec.\ \ref{sec:DGcnf}). $(b)$ The dimer partition function, also including the multivalent terms of Sec.\ \ref{sec:multivalent}, enables to calculate the probability to form dimers.}\label{Fig:ThermInt}
\end{figure}

\subsection{Numerical estimate of configurational effects }
\label{sec:freeenergy}

In this section we explain our strategy to combine the analytical partition functions of interacting multivalent vesicles $\Omega_\dag$ (with $\dag=\mathrm{B}$ or $\dag=\mathrm{B,L}$, see Sec.\ \ref{sec:multivalent}) with the  numerical calculation of the configurational costs due to non selective terms of the Hamiltonian (${\cal H}_\mathrm{ves}$ in Eq.\ \ref{eq:TotHAm}) as estimated  \emph{via} Monte Carlo.
These terms account for the deformation of the membranes following the formation of flat adhesion patches in the interacting vesicles, for the steric repulsion between membranes, as well as for the entropic costs of confining linkers forming bridges within the patch area.  
The role played by elastic and entropic terms in non--selective adhesion of vesicles with different morphologies has been 
 extensively studied in the past.\cite{lipowsky1991adhesion,lipowsky1986unbinding,gruhn2005temperature}
However a direct comparison with previous results is not  straightforward in view of the many peculiarities of multivalent ligand/receptor interactions.\cite{patrick-jcp,hu2013binding}\\
Fig.\ \ref{Fig:ThermInt}~($a$) sketches the thermodynamic path we employ for the simulations. First, as detailed in Sec.\ \ref{sec:Q1}, we use a hit-or-miss algorithm to calculate the internal partition function $z\left(\Omega_\dag,n_\mathrm{B}=1\right)$ of two vesicles linked by a single bridge with respect to the reference state of two free vesicles.
By definition, the internal partition function includes the contributions of all degrees of freedom but the centre of mass of the free/linked vesicles. In Sec.\ \ref{sec:DGcnf} we then use the results of Sec.\ \ref{sec:Q1} to calculate the internal partition function  of vesicles featuring an arbitrary number ($n_\mathrm{B}$) of  bridges $z(\Omega_\dag, n_\mathrm{B})$. In Sec.~\ref{sec:sim} we briefly outline our simulation algorithm.

\subsubsection{Simulation procedure}\label{sec:sim}
In a single simulation cycle we sample on average all degrees of freedom of the system by means of local Monte Carlo moves.
We attempt to randomly displace the nodes of the triangulated membrane, including the linkers possibly present on them, using the Metropolis algorithm and the Hamiltonian given in Eq.~\ref{eq:ves}.  The fluidity of the lipid membrane is simulated by bond-switch Monte Carlo moves described in Refs.~\cite{Ho_PRA,noguchi2009membrane}
Local MC moves result in slow relaxation with autocorrelation times of the slowest fluctuation mode of the order of $t_\mathrm{autcorr}=10^4\,$MC cycles for the systems studied in this work. To enable sufficient sampling all simulated trajectories are longer than $5\cdot 10^5$ MC cycles.
Consistently with the fact of having a fluid lipid membrane, linkers are allowed to diffuse on the triangulated membrane.
For computational efficiency, the diffusion of the linkers on the triangulated mesh has been implemented using random jumps in which a randomly chosen linker is first selected and then moved to a randomly selected free vertex.
All displacements that take two bound linkers to a distance bigger than $L$ are rejected.
The sampling of the bonds is done using a heat-bath algorithm.\cite{miriam}
In particular we randomly choose a linker $i$ and create a list ${\cal L}_i$ of all possible complementary linkers that could potentially form a bond with $i$, eventually including the linker to which $i$ is already bound.
The selected linker has then a probability $p_\mathrm{free}$ of  becoming (or remaining) free and a probability $n({\cal L}_i)  p_\mathrm{bound}$ of getting (or staying) connected to a randomly chosen partner from the list ${\cal L}_i$, which counts $n({\cal L}_i)$ elements. Consistently with Eqs.~\ref{eq:ham-selec} and \ref{eq:TotHAm}, the  probabilities are defined as
\begin{eqnarray}
p_\mathrm{free} &=& {1\over 1+ n({\cal L}_i) \exp [-\beta \Delta G_\mathrm{B}]}
\nonumber \\
p_\mathrm{bound} &=&  { \exp [-\beta \Delta G_\mathrm{B}] \over 1+ n({\cal L}_i) \exp [-\beta \Delta G_\mathrm{B}] } \, \, .
\end{eqnarray}
Note that as explained in Sec.\ \ref{sec:multivalent} for efficiency reasons we simulate explicitly only the formation of bridges.\\
When studying systems of adhering vesicles (Sec.\ \ref{sec:DGcnf}) we also attempt moves in which a  vesicle is randomly chosen and rigidly translated along a random vector. The move is rejected if it causes two vesicles to overlap or a formed bridge to stretch beyond its maximum allowed bond-length $L$.

\subsubsection{Configurational costs of forming the first bridge}
\label{sec:Q1}

In this section we calculate the internal partition function $z(\Omega_\dag,n_B=1)$ of two vesicles bound by a single linkage by taking as a reference the internal partition function of two separate vesicles [first step in Fig.\ \ref{Fig:ThermInt}~($a$)]. 
Such partition function is given by the sum over all allowed conformations in systems of two vesicles, weighted by the number inter--vesicle linkages possible in each conformation. We employ a Monte Carlo algorithm resembling what previously used to calculate the virial expansions of single chain observables of polymer suspensions \cite{li1995critical,caracciolo2008two} or pair interactions between particles functionalised by inert polymer brushes. \cite{mladek2011pair}
We divide the runs into $M=25$ steps. At each step $i$ ($i=1,\cdots , M$) we thermalise two non-interacting  vesicles with $N_\mathrm{therm}=10000$ MC cycles defined in Sec.\ \ref{sec:sim} and use the final configurations to estimate 
the internal partition function $Q_1$ of two vesicles each featuring a single linker.
In particular we choose $N_\mathrm{tries}=250$ random displacements between the centres of mass of the two vesicles (${\bf r}_j$, $j=1,\cdots,N_\mathrm{tries}$) uniformly sampled from the spherical shell of inner radius $R_{\mathrm{min},i}$ and outer radius $R_{\mathrm{max},i}$.
The radii $R_{\mathrm{min},i}$ and $R_{\mathrm{max},i}$ are chosen in a way to guarantee that all possible displacements between vesicle's centre of mass resulting in a linked configurations satisfy the condition $R_{\mathrm{min},i}<|{\bf r}|<R_{\mathrm{max},i}$. Additionally a random rotation of the displaced vesicle around its centre of mass is performed.
For each displacement and rotation, we loop over all the pairs of vertices belonging to different vesicles and count the number $n^\mathrm{bonds}_i$ of possible inter-particle bonds, with $n^\mathrm{bonds}_i=0$ if the two vesicles overlap. 
$Q\left(1\right)_{i}$ is then estimated as 
\begin{eqnarray}
Q\left(1\right)_{i} &=& \frac{\sum\limits_{j=1}^{N_\mathrm{tries}}n^\mathrm{bonds}_j}{N_\mathrm{tries}}v_{0,i}\frac{1}{N_\mathrm{v}^2} \, \, , 
  \label{eq:q1_single} \\
v_{0,i} &=& \frac{4\pi}{3}\left(R_{\mathrm{max},i}^3-R_{\mathrm{min},i}^3\right) \, \, , 
\nonumber 
\end{eqnarray}
 where $1/N_\mathrm{v}^2$ accounts for the probability of finding one linker on a given node of the membrane 
 and the volume $v_{0,i}$ encloses the displacement between the vesicles' centres of mass we sample, accounting for both changes in the absolute distance between the vesicles and rotations of one LUV around the other.
The final value of $Q(1)$ is sampled using 
\begin{eqnarray}
 Q\left(1\right) = \frac{1}{N_\mathrm{cycle}} \sum\limits_{i=1}^{N_\mathrm{cycle}} Q\left(1\right)_{i} \, .
  \label{eq:q1_total}
\end{eqnarray}
The internal partition function of two vesicles featuring a given multivalent model $\Omega_\dag$ (see Sec.\ \ref{sec:multivalent})
linked by one bridge is then given by 
\begin{eqnarray}
 z\left(\Omega_\dag, 1\right)=Q\left(1\right)\Omega_\dag\left(N,1\right) \, \, .
 \label{eq:z1}
\end{eqnarray}
The values of $Q(1)$ for the systems considered in this work have been reported in Tab.\ \ref{Tab:systems}.

\subsubsection{ Configurational costs of forming $n_\mathrm{B}$ bridges}
\label{sec:DGcnf}

\begin{figure}[t!]
\includegraphics[width=9cm]{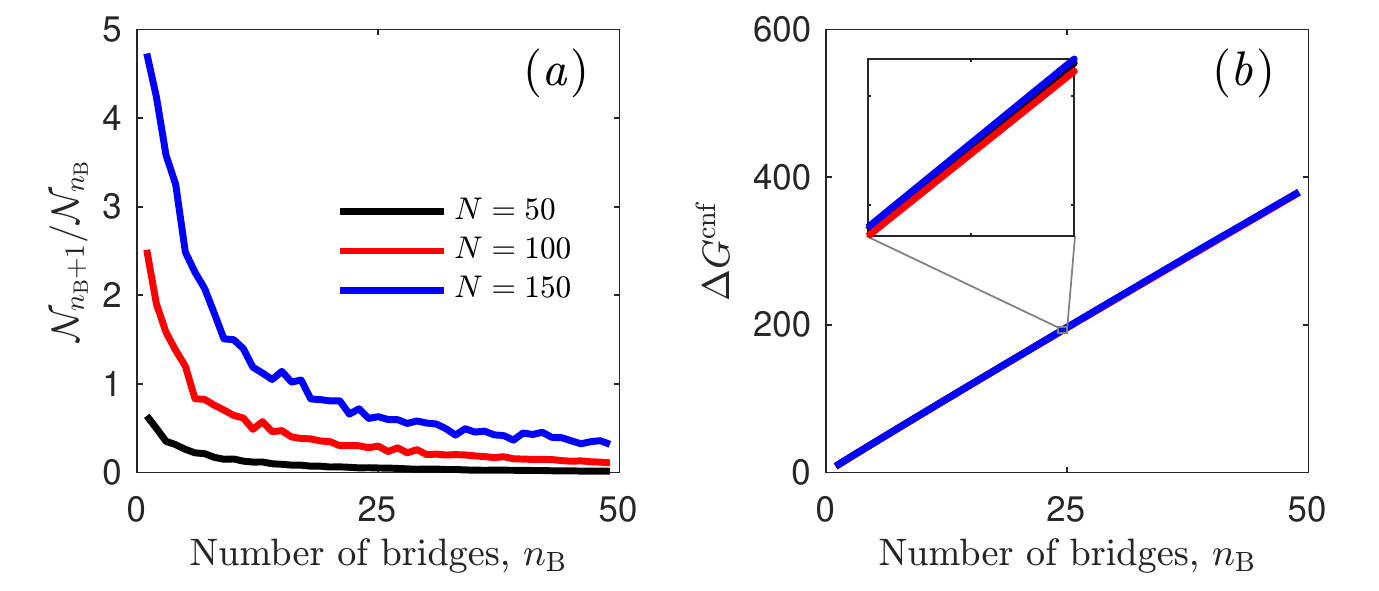} 
\caption{ Evaluating configurational free energy in pairs of interacting vesicles from MC simulations.
$(a)$ Ratio ${{\cal N}_{n_\mathrm{B}+1} / {\cal N}_{n_\mathrm{B}} }$ between the number of times the system visits a configuration with $n_\mathrm{B}+1$ bridges and one with $n_\mathrm{B}$ bridges for different number of linkers (\emph{per} type) $N$. Model parameters: $\Delta G_\mathrm{0}=0$ kJ mol$^{-1}$, $N_\mathrm{v} = 2004$, and $\Omega_\dag=\Omega_\mathrm{B}$. $(b)$ Computed configurational free energy $\DGcnf(1\to n_\mathrm{B})$ as a function of $n_\mathrm{B}$ for the same values of $N$ as in panel ($a$). No significant dependence on $N$ is observed.
Note that accordingly to Eq.\ \ref{eq:deltagconf} a different choice of $\Delta G_\mathrm{0}$ for the calculation of ${{\cal N}_{n_\mathrm{B}+1} / {\cal N}_{n_\mathrm{B}} }$ would produce identical $\DGcnf$.}
\label{fig:figure_3}
\end{figure}

The internal partition function of vesicles linked by $n_\mathrm{B}$ bridges $z(\Omega_\dag ,n_\mathrm{B})$    is sampled using the successive umbrella sampling of Virnau and M\"uller \cite{virnau2004calculation} taking as bias parameter the number of bridges $n_\mathrm{B}$ [second step in Fig.\ \ref{Fig:ThermInt}~($a$)].
We successively constrain the algorithm to sample between states with $n_\mathrm{B}$ and $n_\mathrm{B}+1$ bridges. The ratio between the internal partition functions of the two states can then be directly evaluated as $z(\Omega_\dag,n_\mathrm{B}+1)/z(\Omega_\dag, n_\mathrm{B})= {\cal N}_{n_\mathrm{B}+1}/{\cal N}_{n_\mathrm{B}}$, where ${\cal N}_{n_\mathrm{B}+1}$ and ${\cal N}_{n_\mathrm{B}}$ count the number of times the system visits states with $n_{B}+1$ and $n_\mathrm{B}$ bridges respectively. Using $z(\Omega_\dag, 1)$ as calculated in Sec.\ \ref{sec:Q1} we then obtain 
 \begin{eqnarray}
  z\left(\Omega_\dag, n_\mathrm{B}\right) &=& z\left(\Omega_\dag,1 \right)  
  \prod_{\tau=1}^{n_\mathrm{B}-1} {{\cal N}_{\tau+1} \over {\cal N}_{\tau} }
 \end{eqnarray}
 In Fig.~\ref{fig:figure_3}~(\emph{a})  the ratio ${\cal N}_{n_\mathrm{B}+1}/{\cal N}_{n_\mathrm{B}}$ is shown as a function of $n_\mathrm{B}$ for a bridge--only model ($\dag=\mathrm{B}$) with different amounts of linkers $N$. 
 Similar to what done in Eq.\ \ref{eq:z1}, we factorise the internal partition function using the selective term $\Omega_\dag$ and a configurational term indicated as $\DGcnf$
\begin{eqnarray}
  \frac{z\left(\Omega_\dag, n_\mathrm{B}\right)}{z\left(\Omega_\dag , 1\right)} &=& \exp\left[-\beta \DGcnf(1\to n_\mathrm{B})\right] {\Omega_\dag(N, n_\mathrm{B}) \over \Omega_\dag(N,1)} \, \, ,
\label{eq:partfunct}
\end{eqnarray}
from which we derive 
 \begin{eqnarray}
  \beta \DGcnf(1\to n_\mathrm{B}) &=&- { \sum_{\tau=1}^{n_\mathrm{B}-1} }\log 
  \Big( { {\cal N}_{\tau+1} \over {\cal N}_{\tau} } {\Omega_\dag(N,\tau) \over \Omega_\dag(N,\tau+1)} \Big) \, \, .
  \label{eq:deltagconf}
 \end{eqnarray}
 Note that in view of Eq.\ \ref{eq:partfunct}, $\Delta G^\mathrm{cnf}$ is not a function the hybridisation free energy of single linkers ($\Delta G_B$ and $\Delta G_L$ in Eq.\ \ref{eq:deltagl}).
The contribution $\DGcnf$ describes all the non-specific configurational effects that are not already included in the analytical multivalent partition function. Specifically, it accounts for membranes configurational entropy, and steric repulsion between the membranes, and confinement of the bridge-forming linkers to within the formed adhesion patch.
For ideal linkers that do not interact sterically with each other or with the vesicles, $ \DGcnf$ is only a function of the number of bridges $n_\mathrm{B}$, while it does not depend on the number of linkers $N$ or the particular multivalent model used (as specified by $\dag$ in $\Omega_\dag(n_\mathrm{B},\tau))$.
This has been verified in Fig.\ \ref{fig:figure_3}~($b$) where we calculated $\DGcnf(1\to n_\mathrm{B})$ by using a bridge-only model ($\dag=\mathrm{B}$, see Sec.\ \ref{sec:multivalent}) and three different values of $N$ ($N=50, \,100,\, 150$).
The results show an universal trend for $\DGcnf$. We observe that $\DGcnf$ increases almost linearly with $n_\mathrm{B}$, which enables the extrapolation of  $\DGcnf$ to values of $n_\mathrm{B}$ above the simulated range.

% %% %%%%%%%%%%%%%%%%%%%%%%%%%%%%%%%%%%%%
% %% %%%% FIGURE 3  %%%%%%%%%%%%%%%%%%%%%%%%%%
% %% %%%%%%%%%%%%%%%%%%%%%%%%%%%%%%%%%%%%

\subsection{Dimerisation probability}
\label{sec:Pdim}

To study the melting transition of the DNA-functionalised veiscles we calculate the probability of forming liposome dimers starting from a dilute suspension of free vesicles [see Fig. \ref{Fig:ThermInt}~($b$)]. We start by evaluating the internal partition function of pairs of bound vesicles ($Z$) that is obtained by summing $z(\Omega_\dag,n_\mathrm{B})$ over all possible number of bridges
\begin{eqnarray}
   Z=\frac{1}{\eta} \sum\limits_{n_\mathrm{B}=1}^{N} z(n_\mathrm{B},\Omega_\dag) 
   \label{eq:Z_tot}
\end{eqnarray}
Here $\eta$ is symmetry factor equal to 2 if the vesicles are identical, to 1 otherwise.
The dimerisation probability of two vesicles in a solution with a specific vesicle concentration $c^\mathrm{exp}$ can be written as
\begin{eqnarray}
P_{\text{dim}} = (1 - P_{\text{dim}})^2Zc^\mathrm{exp}N_{A},
\end{eqnarray}
 which can be solved for $P_{\text{dim}}$ leading to the final expression
\begin{eqnarray}
 P_{\text{dim}}=\frac{2Zc^\mathrm{exp}N_{A}+1-\sqrt{4Zc^\mathrm{exp} N_{A}+1}}{2Zc^\mathrm{exp} N_{A}} \, \, .
 \label{eq:Pdim}
\end{eqnarray}
 Note that concentrations are here expressed in units of mol m$^{-3}$, and $N_\mathrm{A}$ is Avogadro's number.\\
It is important to stress the portability of our method. In view of the universality of $\DGcnf$ [Eq.\ \ref{eq:deltagconf} and Fig.\ \ref{fig:figure_3}($b$)], our procedure allows to calculate the internal partition function in pairs of vesicles at different temperatures or number of linkers $N$ by re-weighting $\DGcnf$ with the proper multivalent partition function (see Eq.\ \ref{eq:partfunct}), without the need of further MC simulations.

\subsection{ Other quantified observables }\label{Sec:Analysis}
Besides the configurational contribution to the vesicle-vesicle interaction free energy $\DGcnf$, and the vesicle dimerisation probability $P_{\text{dim}}$, we quantify  the area $A_\mathrm{p}$ of the adhesion patch, and the number of formed intra-vesicle bridges and inter-vesicle loops.\\
 The adhesion patch between vesicles [see Fig. \ref{fig:figure1}~($b$)] is defined as the region featuring vertices that could 
 potentially bind to at least one vertex on the partner vesicle. Note that the exact position of the patch border depends on $L$. For numerical efficiency the patch area is once evaluated every 2500 MC 
 steps.
From the area $A_\mathrm{p}$ of the adhesion patch, we can evaluate the confining contribution to the configurational free energy of bridge formation as
\begin{eqnarray}
 \Delta G_\mathrm{B}^\mathrm{confining} = - k_B T \log\left(\frac{{ L^2} A_\mathrm{p}}{A^{2}}\right) 
\label{eq:confiningB}
\end{eqnarray}
where $A$ the total area of the vesicle. As discussed in Sec.~\ref{sec:multivalent}, $\Delta G_\mathrm{B}^\mathrm{confining}$ describes the loss of translational freedom following the formation of a bridge, when two initially free linkers become confined to a distance $L$ from each other and within the adhesion patch. \\
The average number of bridges  is calculated as
\begin{eqnarray}
 \left\langle n_{B} \right\rangle = \frac{\sum\limits_{n_\mathrm{B}=1}^{N}z\left( \Omega_\dag, n_\mathrm{B}\right)n_\mathrm{B}}{\sum\limits_{n_\mathrm{B}=1}^{N}z\left(\Omega_\dag , n_\mathrm{B}\right)}
 \label{eq:average_bridge}
\end{eqnarray}
If loops are present, their average number can be estimated using a saddle point approximation of the multivalent partition function in Eq.~\ref{eq:multivalency-l}, resulting in
\begin{eqnarray}
\on_L\left(N,n_\mathrm{B}\right) = (N- {n_\mathrm{B}\over 2} - \on_L)^2 e^{-\beta \Delta G_\mathrm{L}}
\label{eq:loops}
\end{eqnarray}
Using this expression along with Eq.\ \ref{eq:average_bridge} the average amount of  bound linkers (loops + bridges) in pairs of connected vesicles is given by
\begin{eqnarray}
\left\langle n_{B+L} \right\rangle_\mathrm{dim}= \frac{\sum\limits_{n_\mathrm{B}=1}^{N}z\left(n_\mathrm{B},\Omega_\dag\right)\left(n_\mathrm{B}+2\on_L(N,n_\mathrm{B}) \right)}{\sum\limits_{n_\mathrm{B}=1}^{N}z\left(n_\mathrm{B},\Omega_\dag\right)}
\end{eqnarray}
Loops can also form in pairs of vesicles that are not connected. Accounting for this possibility we can estimate the total number of bound linkers as
\begin{eqnarray}
 \left\langle n_{B+L} \right\rangle_{tot} = P_{\text{dim}}\left\langle n_{B+L} \right\rangle_\mathrm{dim}+\left(1-P_{\text{dim}}\right)2\on_L
 \label{eq:bridgeandloop}
\end{eqnarray}
where $P_{\text{dim}}$ is given in Eq.\ \ref{eq:Pdim}.

 Vesicle dimerisation temperature is estimated by evaluating  the temperature dependence of $P_{\text{dim}}$. Starting from $T=0\,^\circ$C and incrementing $T$ in steps of 1$\,^\circ$C, the dimerisation temperature is defined as the first point where $P_{\text{dim}}$ drops below 0.5.\\
Similarly, the melting temperature of the DNA linkers is defined as the temperature above which less than 50\% of the linkers are hybridised.

\subsection{Simulated Systems}
\label{sec:ana}

\begin{table}
\small
  \caption{List of simulated systems featuring different vesicle size (number of vertices in the mesh) $N_\mathrm{v}$, maximum linker length $L$ and bending rigidity $\kappa$. In the fourth column we report the partition function of a pair of vesicles linked by a single bridge (Eq.~\ref{eq:q1_total}). Relative statistical errors are $\approx$5-10\%. }
  \label{Tab:systems}
  \begin{tabular*}{0.48\textwidth}{@{\extracolsep{\fill}}cccc}
    \hline
    Vertices $N_v$ & Linker Length $L/\sigma$ & Bending rigidity $\kappa / k_\mathrm{B}T$ & $Q_1 / \sigma^3$ \\
    \hline
    8004 & 3 & 20 & 0.15\\
    4504 & 3 & 20 & 0.22\\
    2004 & 3 & 20 & 0.40\\
    4504 & 5 & 20 & 2.98\\
    4504 & 10 & 20 & 67.14\\
    4504 & 3 & 5 & 0.16\\
    4504 & 3 & 30 & 0.23\\
    \hline
  \end{tabular*}
\end{table}

Table \ref{Tab:systems} lists the parameters of the simulated systems.
We explored the influence of vesicle size, the maximum distance between bound linkers $L$, and the bending rigidity of the membrane $\kappa $ on different properties of interest. Simulation units are converted to physical units by comparing the diameter of the vesicles in experiments $D^\mathrm{exp}$, with that of the simulated liposomes $D^\mathrm{sim}$. 
In particular the relation $D^\mathrm{sim}\sigma=D^\mathrm{exp}$ allows to convert  the simulation unit length $\sigma$ and the linkers' binding range $L^\mathrm{exp}=L \sigma$ in physical units. In the system with $N_\mathrm{v} = 8004$ and $L=3\sigma$, using $D^\mathrm{exp}\approx 400$nm we find $\sigma = 9.1$nm.\\

%%%%%%%%%%%%%%%%%%%%%%%%%%%%%%%%%%%%%%%%%%
%%%%%%%%%% RESULTS %%%%%%%%%%%%%%%%%%%%%%%%%%
%%%%%%%%%%%%%%%%%%%%%%%%%%%%%%%%%%%%%%%%%%

\section{Results and discussion}
\label{sec:res}

\subsection{Configurational free energy}

In this section we discuss numerical estimates of the configurational contribution $\DGcnf(1\to{n_\mathrm{B}})$ to the interaction free energy of pairs of liposomes linked by $n_\mathrm{B}$ bridges, and disentangle components deriving from membrane deformation and steric repulsion from those caused by the confinement of bridge-forming linkers within the inter-vesicle adhesion patch.
Figure~\ref{fig:figure4} shows the configurational free energy $\DGcnf(1\to{n_\mathrm{B}})$ and the area $A_\mathrm{p}$ of the inter-vesicle adhesion patch as a function of the number of formed bridges  at different vesicle size [($a$) and ($b$)], linker length $L$  [($c$) and ($d$)], and membrane bending modulus $\kappa$  [($e$) and ($f$)].
In all cases, we  find that $\DGcnf$ increases linearly with the number of bridges, and this is correlated with the formation of larger adhesion patches
 at higher $n_\mathrm{B}$. 
However the patch area is non--linear in $n_\mathrm{B}$ with $A_\mathrm{p}$ that increases more rapidly
for small number of bridges. 
 Similar trends have been previously reported 
 when studying adhesion between fluid vesicles and solid supports at different vesicle--substrate attraction. \cite{gruhn2005temperature,linke2007adhesion}\\
% However the patch area  increases sub-linearly with $n_B$. Similar trends have been previously reported 
% when studying adhesion between fluid vesicles and solid supports at different vesicle--substrate attraction. \cite{gruhn2005temperature,linke2007adhesion}\\
As discussed in Sec. \ref{sec:multivalent}, $\DGcnf$ implicitly includes the entropic cost $\Delta G_\mathrm{B}^\mathrm{confining}$ of confining each linker engaged in a bridge within the patch region. This term can be estimated using the measured patch area (right column of Fig. \ref{fig:figure4}) and Eq.~\ref{eq:confiningB}, and it is shown in Fig. \ref{fig:figure4} (left column, circles). 
Interestingly,  $n_\mathrm{B} \Delta G_\mathrm{B}^\mathrm{confining}$ accounts for the entire configurational free-energy costs of binding a pair of vesicles, regardless on the simulation conditions tested in Fig.~\ref{fig:figure4}. Deviations of $n_\mathrm{B} \Delta G_\mathrm{B}^\mathrm{confining}$ from $\DGcnf$ are difficult to quantify due to uncertainties in the estimation of $A_\mathrm{p}$. From this observation we deduce that translational entropic terms hindering bridge formation are largely dominant over other contributions to $\DGcnf$. In particular steric repulsion between the membranes, and membrane stretching following adhesion have negligible effect. Note that $\DGcnf$ accounts for the overall configurational free energy fo the system, while $\Delta G_\mathrm{B}^\mathrm{confining}$ as derived in Eq.~\ref{eq:confiningB} accounts for the contribution of a single linker, hence we compare $\DGcnf$ with $n_\mathrm{B} \Delta G_\mathrm{B}^\mathrm{confining}$. \\
 Nonetheless, changing physical parameters of the vesicles causes changes in the area of the adhesion patch, and thereby in $\Delta G_\mathrm{B}^\mathrm{confining}$ and $\DGcnf$. In Fig.~\ref{fig:figure4}~($a$)-($b$) we explore the effect of changing vesicle size. The patch area increases with the number of vertices $N_\mathrm{v}$, but the ratio $L^2 A_\mathrm{p} / A^2$ decreases, causing the configurational penalty for bridge formation to become more severe, effectively weakening the attraction between membranes.
\\
Similar considerations can be used to understand the results of Fig.~\ref{fig:figure4}~($c$)-($d$) where we study the effect of changing the bond length $L$.
Increasing $L$ results in more stable bridges (see Eq.~\ref{eq:confiningB}) and therefore in larger adhesion patches.
However, especially for large $L$, the increase in patch area is also due to the fact that bridges made by longer tethers can explore a wider portion of the curved membrane region at the periphery of the adhesion patch. 
In Fig. ~\ref{fig:figure4}~($e$)-($f$) we test the effect of changing membrane bending modulus $\kappa$. In the range of values we tested, centred around the experimental bending modulus of DOPC bilayers $\kappa \approx 19 k_\mathrm{B}T$,~\cite{Rawicz2000,Sorre2009,Rautu_arXiv} 
we observe little effect on the patch size and thereby on the configurational free energy.
This evidence confirms that contributions arising from the elastic deformation and the suppression of membrane thermal fluctuations are overwhelmed by the entropic terms related to bridge formation. \cite{lipowsky1991adhesion,lipowsky1986unbinding} Our findings are consistent with the observation that the effect of thermal fluctuations becomes negligible for liposomes in the size-range of LUVs.\cite{seifert1990adhesion}\\

%%%%%%%%%%%%%%%%%%%%%%%%%%%%%%%%%%%%%%%%%%
%%%%%%%%%    FIGURE 4     %%%%%%%%%%%%%%%%%%%%%%%%%
%%%%%%%%%%%%%%%%%%%%%%%%%%%%%%%%%%%%%%%%%%

\begin{figure}[ht!]
\includegraphics[width=9cm]{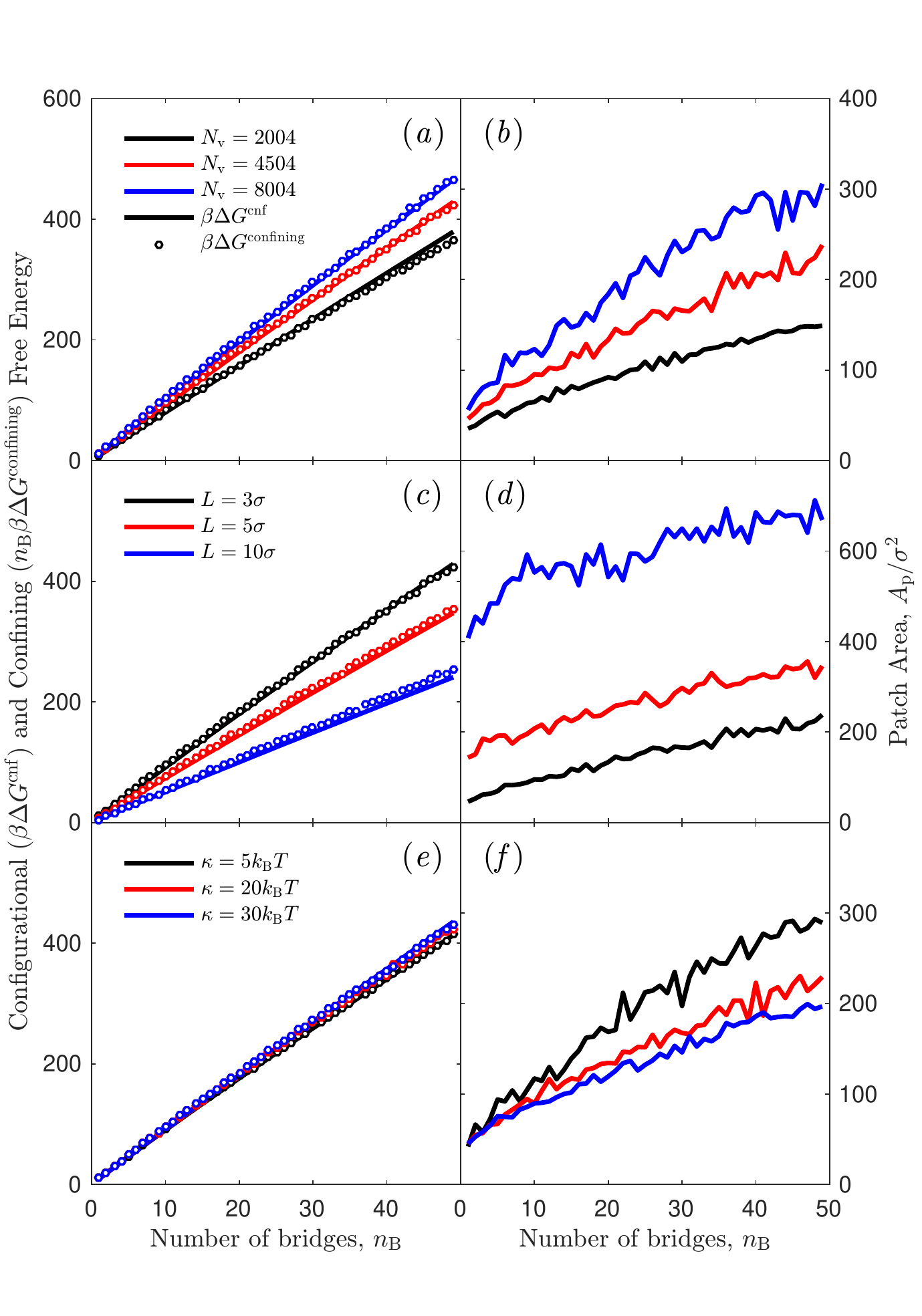} 
\caption{\label{fig:figure4}
Simulated configurational free energy for the formation of vesicle dimers $\DGcnf(1\to n_\mathrm{B})$  (left column) and area of the adhesion patch $A_\mathrm{p}$ (right column) as a function of the number of inter-vesicle bridges $n_\mathrm{B}$. ($a$)-($b$) Effect of changing vesicle size (number of vertices in the triangulated mesh) $N_\mathrm{v}$. ($c$)-($d$) Effect of changing maximum bond length $L$. ($e$)-($f$) Effect of changing bending modulus $\kappa$. Circles in the left column mark the confinement contribution $\Delta G_\mathrm{B}^\mathrm{confinement}$ to $\DGcnf(1\to n_\mathrm{B})$ evaluated using Eq.~\ref{eq:confiningB}  (multiplied by the number of bridges $n_B$) and values of $A_\mathrm{p}$ shown in the right column.
}\label{fig:figure4}
\end{figure}

 \subsection{Vesicle aggregation and DNA hybridisation}
\label{sec:DiscMelting}

\begin{figure*}[ht!]
\includegraphics[width=19.5cm]{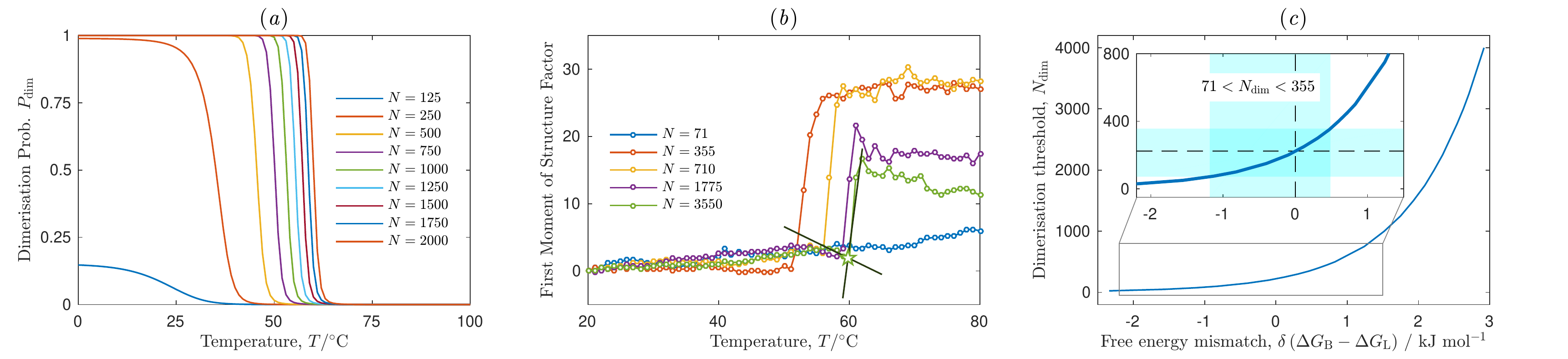} 
\caption{Temperature-dependent dimerisation of vesicles. ($a$) Simulated dimerisation probability as a function of temperature and number of linkers (\emph{per} type) $N$. Model parameters: $N_\mathrm{v} = 8004$, $L=3\sigma$. ($b$) Experimental first moment of the structure factor in aggregating LUVs as a function of temperature and $N$. Data are recorded while heating up aggregated samples. Straight lines (solid) are used to fit the curves just below and just above the melting transition, and identify the melting temperature (star symbol). Samples with $N= {71}$ never show aggregation, while samples with $N={355}$ always aggregate. Samples with $N={177}$ aggregate in 2 out of 5 independent repetitions  of the experiment (not shown for clarity), indicating that $N={177}$ is close to the threshold value for vesicle dimerisation $N_\mathrm{dim}$.
First moment curves collected on cooling show a smoother trend due to the slow aggregation kinetics, and are unsuitable to accurately locate the phase transition. If multiple cooling/heating ramps are performed, we observe a systematic shift of the melting temperature towards lower $T$. We ascribe this effect to vesicle degradation resulting for the repeated exposure to high temperatures. ($c$) Predicted $N_\mathrm{dim}$ as a function of an hypothetical unbalance $\delta \left(\Delta G_\mathrm{B} - \Delta G_\mathrm{L}\right)$ between bridge and loop-formation free energies, where $\delta \left(\Delta G_\mathrm{B} - \Delta G_\mathrm{L}\right) = 0$ marks our original choice. In the inset  the shaded region highlights the experimentally determined range for the threshold value ${71}<N_\mathrm{dim}<{355}$.
\label{fig:figure5}
}
\end{figure*}

In this section we study  temperature dependent self-assembly of vesicle suspensions by means of our model and experiments. We show that experiments and modelling agree on the fact that at low number of binders $N$ self-assembly is suppressed. 
Numerically, self-assembly is characterised by means of the vesicle dimerisation probability $P_{\text{dim}}$ (see Sec.\ \ref{sec:Pdim}).
In experiments, we make use of the fully automated microscopy/fluorimetry platform described in Sec.~\ref{sec:thermal} and characterise aggregation of vesicles through Fourier analysis of epifluorescence images and the binding/unbinding state of DNA linkers \emph{via} FRET.\\
For a meaningful comparison with experimental data we choose model parameters that better match the experimental ratio between vesicle diameter and linker length, in particular  $N_\mathrm{v}=8004$ $L=3\ \sigma$ and $\kappa =20$ (see Sec.\ \ref{sec:ana}).
As done in experiments (see Fig. \ref{fig:figure1}) we consider identical vesicles functionalised by two types of complementary linkers featuring intra--vesicle loops and inter-vesicle bridges ($\dag=\mathrm{L,B}$ in Sec. \ref{sec:multivalent} and Sec. \ref{sec:DGcnf}).
Figure \ref{fig:figure5}~(\emph{a}) shows the calculated dimerisation probability as a function of temperature and the number of linkers \emph{per} vesicle and type $N$.  
The $P_\mathrm{dim}(T)$ curves describe a sigmoidal shape, shifting towards higher temperature and becoming more sharp as the number of available linkers increases, a characteristic behaviour of multivalent interactions already observed in DNA functionalsied solid particles.\cite{melting-theory2} %Surprisingly, 
As $N$ is decreased,  $P_\mathrm{dim}(T)$ tends to converge to a low-temperature plateau smaller than 1, and eventually smaller than 0.5 for $N=125$, effectively suppressing dimerisation. This is a unique characteristic of multivalent interactions featuring competition between loops and bridges,\cite{stefano-nature,rogers-manoharan} also confirmed by our experiments.
Figure \ref{fig:figure5}~(\emph{b}) shows the experimental curves of the first moment of the structure factor as measured upon heating up vesicle samples with different $N$ from low to high temperature. The step-like features mark the sharp melting of the aggregates. Also in experiments we observe a threshold value of $N$ below which aggregation is suppressed. In samples with $N=177$ partial self-assembly is observed in 2 out of 5 nominally identical repetitions, while the other 3 samples showed no aggregation, indicating that $N={177}$ is close to the threshold value for vesicle clustering. All samples with $N={71}$ showed no sign of aggregation, while all samples with $N={355}$ aggregated.
The presence of a threshold value $N_\mathrm{dim}$ in the number of linkers, below which dimerisation does not take place, can be rationalised by observing that at low temperature, where the overall number of bonds (loops or bridges) is maximised, the driving force for vesicle adhesion is purely entropic and due the combinatorial advantage of having a fraction of the linkers forming bridges.\cite{stefano-nature,rogers-manoharan}   At low $N$ such gain is not sufficient to overcome the repulsive configurational contributions to the free energy, thus vesicle adhesion becomes unfavourable.
The agreement between the experimental and simulated value of $N_\mathrm{dim}$ is remarkable, particularly in view of the high sensitivity of this value to changes in the model parameters. To exemplify such sensitivity, in Fig.\ \ref{fig:figure5}~($c$) we report the simulated $N_\mathrm{dim}$  as a function of a hypothetical unbalance $\delta \left (\Delta G_\mathrm{B} - \Delta G_\mathrm{L}\right)$ between the hybridisation free energy of forming loops and bridges (see Eq. \ref{eq:deltagl}),  where $\delta \left (\Delta G_\mathrm{B} - \Delta G_\mathrm{L}\right) = 0$ marks our original choice described in Sec.~\ref{sec:model}.
 A small bias of 3 k J mol$^{-1}$ between the two free energies produces a theoretical threshold value of $N_\mathrm{dim}\sim 4000$, more than one order of magnitude larger as than the experimental value. In the inset of Fig.\ \ref{fig:figure5}~($c$) we highlight how the experimentally determined $N_\mathrm{dim}$ can only be captured by simulations for uncertainties in the estimated $\Delta G_\mathrm{B} - \Delta G_\mathrm{L}$ smaller than $\pm$ 1  k J mol$^{-1}$. Note that an unbalance between $\Delta G_\mathrm{B}$ and $\Delta G_\mathrm{L}$ would also result from a wrong estimate of $\DGcnf$. In this respect Fig. \ref{fig:figure5} ($c$) certifies the accuracy of our model in conditions of low $N$.

%%%%%%%%%%%%%%%%%%%%%%%%%%%%%%%%%%%%%%%%%%
%%%%%%%%%% FIGURE 5 %%% %%%%%%%%%%%%%%%%%%%%%%%
%%%%%%%%%%%%%%%%%%%%%%%%%%%%%%%%%%%%%%%%%%

\begin{figure*}
\includegraphics[width=19.5cm]{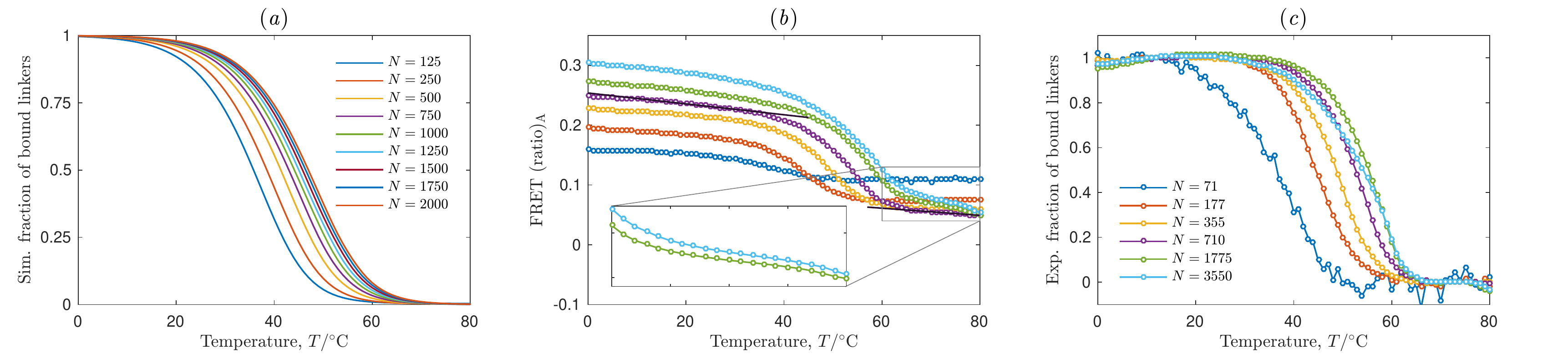} 
\caption{Melting of DNA links. ($a$) Simulated fraction of formed DNA links (loops + bridges) as a function of temperature for systems with variable number of linkers per vesicle. Model parameters: $N_\mathrm{v} = 8004$, $L=3\sigma$.  ($b$) Experimental FRET (ratio)$_\mathrm{A}$ as a function of temperature. Solid lines indicate high- and low-temperature baselines fitted by straight lines (see Sec.~\ref{sec:imageanalysis}). In the inset: a detail of the high temperature plateau in samples with high $N$ showing a second drop in FRET efficiency drop at $\sim 80^{\circ}$C that marks the disassembly of the dsDNA spacers connecting the linkers to the membranes (see Sec.~\ref{sec:DiscMelting}). The curves are obtained averaging over 2 consecutive temperature ramps collected on heating and cooling. No hysteresis is observed.  At high temperature ${\cal I}_\mathrm{Cy3-Cy5}$ (see Eq.~\ref{ratioA}) becomes very small in samples with low DNA concentration and its detection is affected by background noise, \emph{e.g} sample autofluorecence. This causes the high temperature value of (ratio)$_\mathrm{A}$ to increase for samples with low $N$. $c$) Experimental fraction of formed DNA bonds extracted from the ramps of panel ($b$) as detailed in Sec.~\ref{sec:imageanalysis}. Legend applies also to panel ($b$).
}
\label{fig:figure6} 
\end{figure*}

%{\cm
%The novelty of our experimental methodology is the possibility to directly correlate self-assembly  with measures of the number of linkages formed.  
%}
In Fig.~\ref{fig:figure6} we study the fraction of DNA strands engaged either in loops or bridges as a function of temperature, 
experimentally measured on the same systems of Fig. \ref{fig:figure5}.
Panel \emph($a$) reports simulation results obtained using Eq. \ref{eq:bridgeandloop}. Panel (\emph{b}) shows the experimental FRET (ratio)$_\mathrm{A}$, linearly dependent on the FRET efficiency between the fluorophores on complementary linkers. The sigmoidal decrease of FRET efficiency marks the melting of the DNA bonds, while the low and high-temperature plateaux represent the regimes where all the linkers are bound or free, respectively. The fraction of hybridised linkers can therefore be easily estimated by fitting such plateaux with linear baselines as explained in Sec.~\ref{sec:imageanalysis} and shown in Fig.~\ref{fig:figure6}~($b$).\cite{shimobayashi2015direct}  Note however that for high DNA coverage ($N=3555, 1775$), a second small drop in FRET efficiency is observed at high T ($\sim80^\circ$C) as highlighted in the inset of  in Fig.~\ref{fig:figure6}~($b$). In these samples DNA linkers are densely packed, and the average distance between them is comparable to the F\"{o}rster radius of the Cy3-Cy5 pair, causing a non-zero chance of energy transfer also between unbound linkers. When the temperature is increased to $\sim 80^\circ$C, the dsDNA spacers melt and the single-strands carrying the sticky ends and the fluorescent labels, no longer bound to the cholesterol anchors, are released in solution. The detachment of  fluorophore-carrying DNA from the membranes causes the suppression of the small FRET signal ascribed to high-density coating. In samples where the double-transition is present the high-temperature baseline is chosen to fit the plateau observed before the final FRET-efficiency drop, which  particularly for $N=3555$ spans a relatively small temperature range, possibly leading to uncertainty in baseline determination. Similarly uncertainties are found in the samples with the lowest DNA coverage ($N={71}$) and thereby the lowest melting temperature, where the low-temperature plateau spans a small temperature range [see Fig.~\ref{fig:figure6}~($b$)]. The fraction of hybridised DNA linkers as extracted according to this procedure is shown in Fig.~\ref{fig:figure6}~($c$) as a function of temperature and $N$. The DNA melting temperature is extracted as the point where the fraction of hybridised DNA is equal to 0.5, determined \emph{via} linear interpolation.\\
Both simulation and experimental results show the broad melting transition typical of short oligomers.

%%%%%%%%%%%%%%%%%%%%%%%%%%%%%%%%%%%%%%%%%%
%%%%%%%%%% FIGRURE 6 %%%%%%%%%%%%%%%%%%%%%%%
%%%%%%%%%%%%%%%%%%%%%%%%%%%%%%%%%%%%%%%%%%

\begin{figure}[ht!]
\includegraphics[width=9cm]{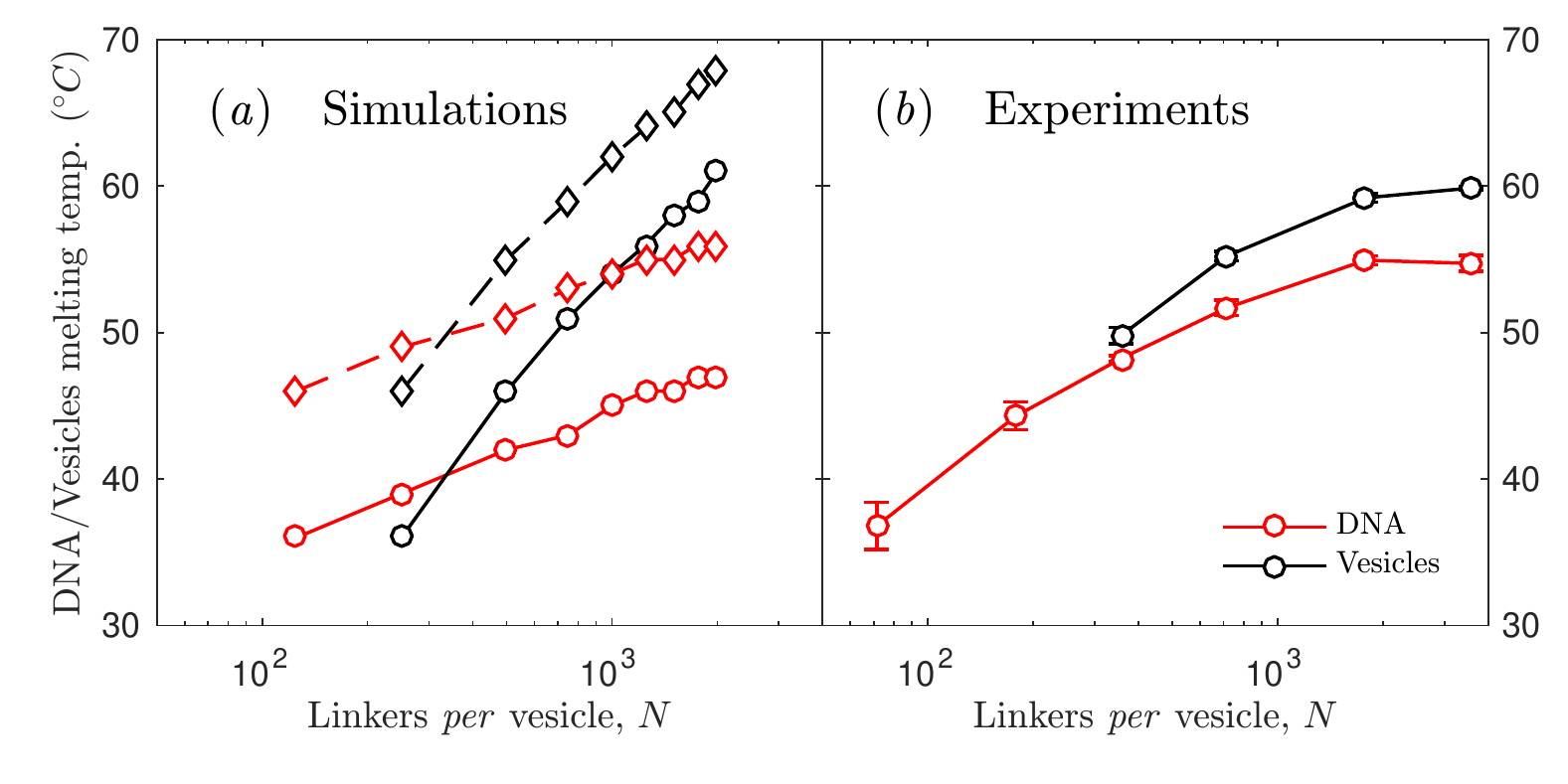} 
\caption{
Melting temperatures for DNA bonds and vesicle aggregation. ($a$) Simulated melting temperatures as a function of the number of linkers \emph{per} vesicle (\emph{per} linker type) $N$. Data shown and solid lines are obtained using hybridisation free energy between DNA sticky ends $\Delta G^0$ evaluated using nearest-neighbours thermodynamic rules on the sequences shown in Fig.~\ref{fig:figure1}~($a$).\cite{santalucia,santalucia2004thermodynamics,dirks2007thermodynamic} To produce dashed curves $\Delta G^0$ has been corrected with an attractive term (2 kJ mol$^{-1}$) to account for the stabilising effect of Cy3 and Cy5 fluorophores.\cite{moreira2015cy3} ($b$) Experimental melting temperatures obtained by averaging over 4 independent experiments. Errorbars indicate standard errors.}
\label{fig:figure7}
\end{figure}

Figure~ \ref{fig:figure7} summarises and compares the dimerisation/aggregation temperature of the vesicles and the melting temperature of DNA linkers, as computed by our simulations [panel ($a$)] and measured experimentally [panel ($b$)].
 Simulations predict a linear increase as a function of $N$ for both the DNA melting temperature and the vesicle dimerisation temperature. However, the slope of the two curves is different, with the vesicle dimerisation temperature increasing more sharply than the DNA melting temperature. As a result, a crossover temperature exists, above which the vesicles can dimerise even when less than 50\% of the DNA bonds are formed, and below which more than 50\% of the available linkers are needed for dimerisation. In Fig.~\ref{fig:figure7}($a$) we also investigate the effect of uncertainties in the hybridisation free energy of the sticky ends $\Delta G^0$. For all results presented in this work, $\Delta G^0$ has been calculated using conventional nearest neighbours rules\cite{santalucia,santalucia2004thermodynamics,dirks2007thermodynamic} applied on the sticky-end sequences shown in Fig.~\ref{fig:figure1}~($a$), including also the attractive effect of the dangling bases ($A$s) neighbouring the hybridising duplexes. However this modelling choice is far from being univocal. On the one hand, it has been demonstrated that the presence of inert DNA tails emanating from the duplexes, like the dsDNA spacers in the present system, can have substantial destabilising effect.\cite{parolini2016,lorenzo-jacs,srinivas2013biophysics,wang2016native} On the other hand, Cy3 and Cy5 fluorophores have a stabilising effect, decreasing the hybridisation free energy by $\approx\,$2 kJ mol$^{-1}$.\cite{moreira2015cy3}
 The dashed lines in Fig.~\ref{fig:figure7}~($a$) exemplify the effect of including this attractive free energy contribution to the vesicle dimerisation and DNA melting temperatures. In both cases we observe a shift of about 10$^\circ$C, which demonstrates how sensitive the present results are to uncertainties in the estimation of $\Delta G^0$.\\
 Experimental data in Fig.~\ref{fig:figure7}~($b$) show that the DNA melting temperature and the vesicle aggregation temperature approach linear dependence on $N$ only at low DNA coverage, and tend to plateau at high $N$. We ascribe this trend to the possible saturation of the lipid membranes when very high concentration of DNA linkers is added in solution. Saturation would result in an actual number of DNA linkers per vesicle smaller than the nominal value calculated as described in Sec.~\ref{sec:prep}. High concentration of hydrophobised linkers may also promote the formation of stable DNA-cholesterol micelles, which would decrease the available number of linkers. As they approach the linear regime at low $N$, also the experimental curves for DNA melting temperature and vesicle aggregation temperature have a different slope, with the latter being more steep. The difference in slope is however less pronounced as compared to simulation results, which causes the vesicles aggregation temperature to reach lower values at high $N$. This discrepancy is possibly due to steric repulsion between linkers, neglected in simulations. In particular, the adhesion patch between two vesicles features an even higher linker concentration than the surrounding free-standing membrane due to the recruitment of bridge-forming linkers.\cite{parolini2014thermal} Steric repulsion within the adhesion patch would therefore limit the number of possible bridges and weaken vesicle-vesicle adhesion.

\begin{figure}[ht!]
\includegraphics[width=9cm]{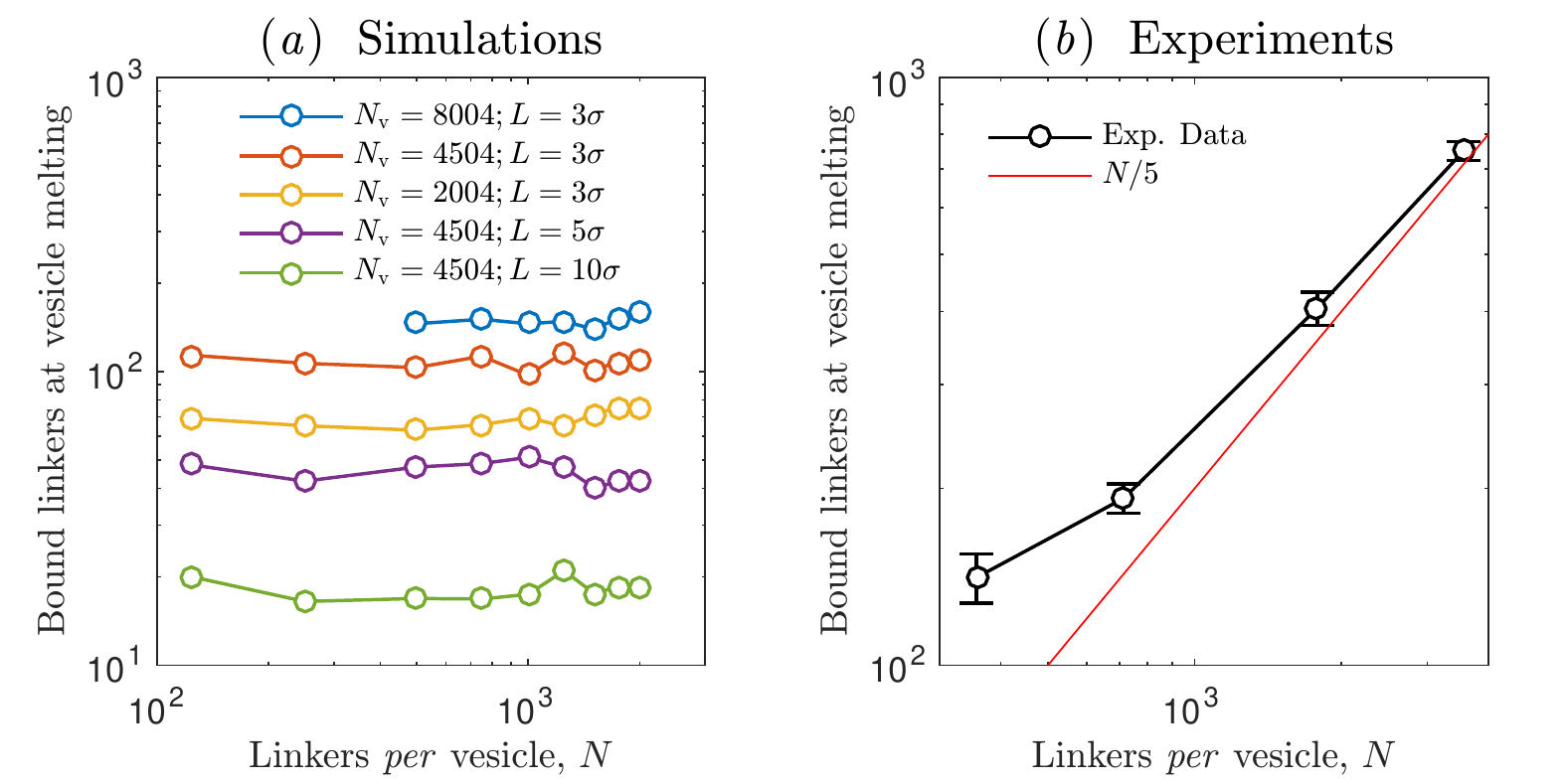} 
\caption{ Fraction of formed DNA bonds evaluated at the melting temperature of vesicle aggregates as a function of the number of DNA linkers per vesicle $N$. ($a$) Simulation results. Curves display a constant trend in a broad range of model parameters ($\kappa = 20 k_\mathrm{B}T$). ($b$) Experimental results. The number of formed DNA bonds at vesicle melting temperature is extracted by linear interpolation of curves of the type shown in Fig.~\ref{fig:figure6}~($c$). Data are averaged over 4 independent experiments. Errorbars indicate standard errors. The curve shows an increasing trend as a function of $N$, approaching a plateau at low $N$ and a linear asymptote ($N/5$) at high $N$.}\label{fig:figure8} 
\end{figure}

A more insightful view in this effect is given by  Fig. \ref{fig:figure8} where we compare the number of formed DNA bonds,  including both loops and bridges, evaluated at the vesicle melting temperature. Simulation results [panel ($a$)] predict that the number of DNA bonds at vesicle melting does not depend on $N$ for a broad range of model parameters.  If steric interactions between the linkers are neglected, multivalent theories predict that the ratio between loops and bridges is independent on temperature or the total number of linkers, and determined only by vesicle geometry.\cite{parolini2014thermal} 
Thus the constant trends in Fig.~\ref{fig:figure8}~($a$) imply that a fixed number of bridges is required to overcome entropic repulsion and bind vesicles to each other.\\
Experiments predict an increase of the fraction of hybridised DNA at vesicle melting as a function of $N$, seemingly approaching a plateau at low $N$ (compatibly with simulations) and a linear asymptote at high $N$. Since it is reasonable to argue that a fixed number of bridges is required to drive vesicle aggregation, we speculate that the observed trend may be caused by a dependence of the bridge/loop ratio on $N$. This could be again ascribed to steric repulsion between linkers within the crowded adhesion patch: Although at high linker concentration the number of formed DNA bonds at vesicle melting is higher, excluded volume effects between linkers in the patch would result in a smaller fraction of bridges and a higher fraction of loops with respect to the ideal scenario in which linker-linker steric interactions are negligible. 

%%%%%%%%%%%%%%%%%%%%%%%%%%%%%%%%%%%%%%%%%%
%%%%%%%%%% FIGURE 7  %%%%%%%%%%%%%%%%%%%%%%%%%%
%%%%%%%%%%%%%%%%%%%%%%%%%%%%%%%%%%%%%%%%%%

%%%%%%%%%%%%%%%%%%%%%%%%%%%%%%%%%%%%%%%%%%
%%%%%%%%%% FIGURE 8  %%%%%%%%%%%%%%%%%%%%%%%%%%
%%%%%%%%%%%%%%%%%%%%%%%%%%%%%%%%%%%%%%%%%%

\section{Conclusions}

Multivalent interactions are at the forefront of many applications in nanotechnology and underpin several functional behaviours in biology. 
 A satisfactory understanding of all aspects of the problem is lacking, particularly in biologically relevant scenarios where substrate deformability couples with configurational free energy contributions and steric interactions. A numerical/theoretical framework validated by dedicated experiments and capable of assessing the consequences of such interplay is therefore of pivotal importance for biological and technological applications.
Modelling multivalent interactions requires a significant coarse graining effort to bridge atomistic scales at which non--covalent bonds form with mesoscopic scales at which functionalised objects (vesicles, surfaces, polymers or nanoparticles) interact as a result of many possible reversible supramolecular linkages.
Analytical theories have been developed to calculate the free energy of multivalent interactions.\cite{melting-theory1,melting-theory2,patrick-jcp,stefano-jcp,di2016communication,tito2016communication,xu2016simple,DeGernier2014,shimobayashi2015direct} 
Considering a set of  interacting multivalent objects, these methods attempt to enumerate all possible configurations of the supramolecular network of linkers mediating the interactions.
The success of these theories enabled the development of efficient simulation schemes for characterising the phase behaviour of multivalent objects, coarse-grained as simple particles with effective pair interactions. Analytical multivalent theories, however, neglect non-selective interactions arising from particle deformability or excluded--volume effects between linkers.\\
In this work we addressed some of the limitations of current modelling approaches to multivalent interactions, particularly the impossibility of accounting for deformable particles. Working on a system of soft liposomes functionalsied by linkers made of synthetic DNA, we propose a method that combines state-of-art multivalent theories with Monte Carlo methods.
In particular we used a triangulated model of the lipid bilayer together with free energy calculations to estimate the configurational free energy cost of having a given number of bridges formed between interacting vesicles.  
We clarify how such contributions are mainly due to the entropic penalty of confining bridge-forming linkers within the flat contact area formed between adhering deformable vesicles. We then characterise the response of pair of interacting vesicles to changes in temperature, and in particular the temperature-dependent dimerisation probability and the melting curves of DNA linkers. Simulation results are compared to experiments on DNA functionalised  Large Unilamellar Vesicles, where the temperature-dependent vesicle aggregation state and DNA hybridisation can be independently monitored by fluorescence microscopy and F\"{o}rster Resonant Energy Transfer thanks to a fully automated and programmable setup. Both simulations and experiments confirm that a minimum number of linkers \emph{per} vesicle is required to overcome configurational entropic costs for membrane deformation and produce stable aggregation.
We observe deviations between simulated and experimental trends at high density of the DNA linkers. We argue that this disagreement is caused by excluded volume effects between pairs of linkers and between the linkers and the membranes, neglected in the present contribution to maximise the portability of the model. These deviations deserve future investigation.\\
Our experimental and numerical results highlight the importance of configurational free energy costs arising from the deformability of objects interacting \emph{via} multivalent interactions. This is an ubiquitous scenario in biological contexts, where deformable cells adhere to each other or to the extra-cellular matrix thorough membrane ligand/receptors, but also in bio-nanotechnology and nano-medicine, where multivalent synthetic probes are designed to selectively target cells. The novel numerical approach we developed to reach these conclusions combining state-of-art analytical modelling with Monte Carlo simulations provides a valuable tool for further investigations of specific biological and nano-technological problems including tissue  dynamics, cell sorting, cell-cell and cell-substrate adhesion, tissue scaffolding, and designing multivalent probes for drug and gene delivery.\\

{\bf Acknowledgements} SJB and BMM are supported by the Universit\'e Libre de Bruxelles (ULB). JK, LP, LDM and PC acknowledge support from the EPSRC Programme Grant CAPITALS number EP/J017566/1. LDM acknowledges support from the Oppenheimer Fund, Emmanuel College Cambridge, the Leverhulme Trust and the Isaac Newton Trust through an Early Career Fellowship. 
AS thanks the Human Frontier Science Program and Emmanuel College Cambridge. 
BMM thanks Ignacio Rondini for contributing to the early stages of this work.  
Computational resources have been provided by the Consortium des {\'{E}}uipements de Calcul Intensif (C{\'{E}}CI), funded by the Fonds de la Recherche Scientifique de Belgique (F.R.S.-FNRS) under Grant No. 2.5020.11.
In compliance with the requirements of
EPSRC the data underlying this publication are available for download at \textbf{[link to be added]}.


%\bibliography{biblio}

\end{document}